\documentclass[aps,pra,twocolumn,floatfix,nofootinbib]{revtex4-2}
\usepackage[colorlinks]{hyperref}
\usepackage[utf8]{inputenc}
\usepackage{graphicx}
\usepackage{tabularx}
\usepackage{multirow}
\usepackage{amssymb}
\usepackage{amsmath}
\usepackage{microtype}
\usepackage{xcolor}
\usepackage{dcolumn}
\usepackage{refcount}
\usepackage{natbib}
\DeclareMathAlphabet\mathbfcal{OMS}{cmsy}{b}{n}

\newcommand{\bra}[1]{\ensuremath{\langle #1|}}	
\newcommand{\ket}[1]{\ensuremath{|#1\rangle}}	

\newcommand{\threej}[6]{\ensuremath{\begin{pmatrix}#1&#2&#3\\#4&#5&#6\end{pmatrix}}}	
\newcommand{\sixj}[6]{\ensuremath{\begin{Bmatrix}#1&#2&#3\\#4&#5&#6\end{Bmatrix}}}	
\newcommand{\ninej}[9]{\ensuremath{\begin{Bmatrix}#1&#2&#3\\#4&#5&#6\\#7&#8&#9\end{Bmatrix}}}
\renewcommand{\v}[1]{\ensuremath{\boldsymbol{#1}}}		
\newcommand{\brad}[1]{\ensuremath{\langle #1||}}
\newcommand{\ketd}[1]{\ensuremath{|| #1\rangle}}
\makeatletter
\newcolumntype{d}[1]{D{.}{.}{#1}}
\newcommand{\doublewidetilde}[1]{{%
  \mathpalette\double@widetilde{#1}%
}}
\newcommand{\double@widetilde}[2]{%
  \sbox\z@{$\m@th#1\widetilde{#2}$}%
  \ht\z@=.9\ht\z@
  \widetilde{\box\z@}%
}

\makeatother 

\begin{document}

\begin{abstract}
The Stark-interference technique is commonly used to amplify the feeble parity-violating signal in atomic experiments. As a result, interpretation of these experiments in terms of electroweak observables requires knowledge of the Stark-induced $E1$ transition amplitudes or, equivalently, transition polarizabilities.
While the literature assumes that these transition polarizabilities do not depend on the nuclear spin, here we prove the contrary. The nuclear spin dependence arises due to hyperfine mixing of atomic states and requires a third-order perturbation theory (one hyperfine interaction and two electric-dipole interactions) treatment.
 We demonstrate that the so far neglected {\em tensor} contribution appears in the transition polarizability and present numerical results for the nuclear-spin-dependent corrections to the $6S_{1/2}\rightarrow{7S_{1/2}}$ transition polarizability in $^{133}$Cs. We investigate the effect of these corrections to transition polarizabilities on the extraction of the $^{133}$Cs anapole moment from the  Boulder experiment [Science {\bf 275}, 1759 (1997)]. We also consider their effect on the extraction of the ratio between the scalar and vector transition polarizabilities from the measurements  [Phys.\ Rev.\ A {\bf{55}}, 2 (1997)]. While the corrections are minor at the current level of experimental accuracy, our analysis provides a framework for future experiments.
\end{abstract}

\title{Nuclear-spin-dependent corrections to the transition polarizability in cesium}
\author{D. Xiao}
    \affiliation{Department of Physics, University of Nevada, Reno, 89557, USA} 
	\author{H. B. Tran Tan}
    \affiliation{Department of Physics, University of Nevada, Reno, 89557, USA} 
\author{A.  Derevianko}
\email[]{andrei@unr.edu}
	\affiliation{Department of Physics, University of Nevada, Reno, 89557, USA}
\date{\today}
\maketitle
\section{Introduction}

In 1988, an experiment performed by the Boulder group~\cite{Noecker1988PrecisionTheory} provided the first evidence of the nuclear-spin-dependent parity-non-conserving (PNC) interactions in $^{133}$Cs atom, which later led to the discovery of the $^{133}$Cs nuclear anapole moment~\cite{WooBenCho97}.
However, the extracted nuclear anapole moment~\cite{WooBenCho97} has been found to disagree with the nuclear-physics determination~\cite{HaxWie01}. 
In nuclear physics, to bridge different manifestations of  PNC, theorists operate in terms of the weak meson-nucleon couplings~\cite{DDH1980}. 
The weak meson-nucleon couplings propagate through 
the nuclear structure evaluation of anapole moments~\cite{HaxWie01} and other nuclear processes. Linear combinations of these couplings can be constrained by comparison with available experimental data and theoretical 
estimates within the Standard Model framework.
In particular, the nuclear physics constraints come from the scattering of polarized protons on unpolarized
protons and $^{4}\rm{He}$ targets as well as the emission of circularly polarized photons from $^{18}\rm{F}$ and $^{19}\rm{F}$ nuclei. These constraints form nuclear experimental bands whose intersection yields the nuclear physics determinations of the couplings. However, the bounds derived from the measured Cs anapole moment lie outside this nuclear physics favored region. Our paper is motivated in part by this tension between the nuclear and atomic physics determinations of the weak meson-nucleon couplings.


The anapole moment is extracted from the difference between the
two measured PNC amplitudes $E1_{\rm{PNC}}$ connecting different hyperfine components of the ground $6S_{1/2}$ and the excited $7S_{1/2}$ states in ${}^{133}$Cs. The Boulder results read~\cite{WooBenCho97} 
\begin{equation}\label{Eq:BoulderResults} 
    \frac{\mathrm{Im}(E1_{\mathrm{PNC}})}{\beta}=\begin{cases}
    -1.6349(80)\, \mathrm{mV/cm}\\
    \,\,\,\,{\rm for}\,\,\,6S_{1/2},\,{F_i=4}\rightarrow{7S_{1/2},\,{F_f=3}}\,,\\
    -1.5576(77) \, \mathrm{mV/cm} \\
    \,\,\,\,{\rm for}\,\,\,6S_{1/2},\,{F_i=3}\rightarrow{7S_{1/2},\,{F_f=4}}\,.
    \end{cases}
\end{equation}
Here, $F$ is the grand total angular momentum in Cs formed by adding the nuclear spin $I=7/2$ and the total electronic angular momentum $J$, and 
 $\beta$ is the vector transition polarizability.
A weighted average of the two values in Eq.~\eqref{Eq:BoulderResults} yields the
nuclear-spin-independent electroweak observable (weak charge), while their difference -- the nuclear-spin-dependent quantity (nuclear anapole moment). 

Notice the appearance of the vector transition polarizability $\beta$ in the results~\eqref{Eq:BoulderResults}, as the Boulder group used the Stark-interference technique~\cite{BouBou75}. This technique amplifies the feeble PNC effect by the means of an externally applied DC electric field which opens an additional Stark-induced excitation pathway for the nominally $E1$-forbidden 
$6S_{1/2} \rightarrow 7S_{1/2}$ transition. Then the transition rate acquires a cross-term between the Stark-induced and PNC amplitudes. This interference term flips sign under parity reversals enabling its experimental extraction. 

One of the  assumptions made in the Boulder analysis is that $\beta$ does not depend on the nuclear spin. Contrary to this assumption, here we identify nuclear spin-dependent corrections to the Stark-induced transition amplitudes or, equivalently, to the transition polarizabilities ($\beta$ in particular).
While the effects of our newly-introduced corrections turn out to be negligible at the Boulder experiment's level of accuracy, our analysis provides a framework for future experimental efforts.

The paper is organized as follows. In Sec.~\ref{Sec:Generalization-Stark-induced-transition}, we review the Stark-interference technique and derive the second-order transition polarizabilities. The hyperfine-mediated corrections to the transition polarizabilities are derived in Sec.~\ref{Sec:Third-order-HPI-trans-pol} and numerically evaluated in Sec.~\ref{Sec:Numerical}. Our reanalysis of the Boulder APV experiment~\cite{WooBenCho97}  is given in Sec.~\ref{Sec:ReinterpretationPNC}. We also compute correction to the experimentally extracted ratio of the vector and scalar transition polarizabilities in Sec.~\ref{Sec:beta-alpha-ratio}. While we keep the discussion sufficiently general, all our numerical work refers to the $6S_{1/2} \rightarrow 7S_{1/2}$ transition in $^{133}$Cs.
Unless stated otherwise, atomic units are used throughout.

\section{Generalization of Stark-induced  transition polarizability}
\label{Sec:Generalization-Stark-induced-transition}

We are interested in driving an electric-dipole transition from the an initial state $i$ to a final state $f$. We assume that these states are of the same parity, precluding $E1$ transitions. To open the otherwise forbidden $E1$ pathway, we apply a DC electric field which admixes intermediate states of opposite parity into $i$ and $f$~\cite{BouBou75}. 
The relevant amplitude for the resulting $E1$ transition between such mixed states can be derived in the second order of perturbation theory (see Ref.~\cite{TraXiaDer23-2ndOrderBeta} for a detailed derivation). The two perturbations are the electric dipole interactions with the applied DC and driving laser fields.
The  Stark-induced transition amplitude $A_{i\rightarrow{f}}$ is conventionally expressed in terms of the transition polarizability $a_{i\rightarrow{f}}$ as $A_{i\rightarrow{f}}= a_{i\rightarrow{f}}\mathcal{E}_s \mathcal{E}_L$,
which factors out $\mathcal{E}_s$ and $\mathcal{E}_L$, the static and laser field amplitudes, respectively. The transition polarizability for the transitions between two $S_{1/2}$ states is  conventionally parameterized as~\cite{BouBou75}
\begin{align}\label{eq:std_form}
a_{i\rightarrow{f}}=
\alpha(\hat{\v{\varepsilon}}\cdot {\hat{\v{e}}}) \delta_{F_iF_f}\delta_{M_iM_f}+
 i \beta(\hat{\v{e}} \times \hat{\v{\varepsilon}}  ) \cdot{\bra{f}\v{\sigma}\ket{i}} \, .
\end{align}
Here, the two atomic-structure-dependent quantities $\alpha$ and $\beta$ are the scalar and the vector transition polarizabilities. The unit vectors $\hat{\v{\varepsilon}}$ and $\hat{\v{e}}$ characterize polarizations of the laser and static electric fields, respectively. The states $i$ and $f$ are hyperfine basis states, e.g,
$\ket{i} = \ket{n_i(IJ_i)F_iM_i}$ is a state of grand total angular momentum $F_i$ obtained by the conventional coupling of the total electron angular momentum $J_i$ and the nuclear spin $I_i$, with $M_i$ and $n_i$
being the magnetic and principal quantum numbers.  The matrix element of Pauli matrices $\v{\sigma}$ is understood as involving the angular parts of the wavefunctions. 

Qualitatively, Eq.~\eqref{eq:std_form} is obtained~\cite{TraXiaDer23-2ndOrderBeta} in the second order of perturbation theory by recouping the product of two dipole couplings  $(\v{D} \cdot \hat{\v{\varepsilon}}) (\v{D} \cdot \hat{\v{e}})$ into a sum over the irreducible tensor operators (ITO) containing scalar products of compound tensors\footnote{A scalar product of two rank-$k$ ITOs is understood as 
$
    {P^{(k)}}\cdot{{Q}^{(k)}} = \sum_{q=-k}^{k} (-1)^q P^{(k)}_{q} {Q}^{(k)}_{-q}\,,
$
and a compound ITO of rank $Q$ is defined as 
$
\{P^{(k_1)}\otimes{R^{(k_2)}}\}_{q}^{(Q)}=\sum_{q_1q_2} C^{Qq}_{k_1q_1k_2q_2} P^{(k_1)}_{q_1} R^{(k_2)}_{q_2}\,,
$
where $q_1$ and $q_2$ label the spherical basis components of the ITOs with $C^{Qq}_{k_1q_1k_2q_2}$ being the conventional Clebsch-Gordan coefficients. }
 $(\hat{\v{\varepsilon}}\otimes{\hat{\v{e}}})^{(Q)} \cdot (\v{D}\otimes \v{D})^{(Q)}$. Here, $\v{D}$ is the electron electric dipole moment operator.
Based on the angular selection rules, the rank
$Q$ can accept the values of 0, 1, and 2, corresponding to the scalar, vector, and tensor contributions. Hereto, previous analyses of the $6S_{1/2} \rightarrow 7S_{1/2}$ transition polarizability in Cs have neglected the tensor ($Q=2$) contribution. The reason for this is that the dipole operators involve only electronic degrees of freedom and the matrix element of the rank-2 tensor between the $S_{1/2}$ states vanishes due to the angular selection rules. However, if we account for the hyperfine interaction (HFI), the states involved would need to be characterized by the grand-total angular momentum $F$ and the tensor contribution would no longer vanish since $F=3$ or $4$ for the hyperfine manifolds attached to the $S_{1/2}$ electronic states in $^{133}$Cs. Notice that the inclusion of the HFI requires a third-order perturbation theory treatment and therefore leads to the tensor contribution being suppressed compared to the scalar and vector contributions.

The tensor contribution to Eq.~\eqref{eq:std_form} can be parameterized as
\begin{equation}
a_{i\rightarrow{f}} = ... + \gamma{\bra{f}\{I\otimes{I}\}^{(2)}\ket{i}(\hat{\v{\varepsilon}}\otimes{\hat{\v{e}}})^{(2)}}\,, \label{Eq:GammaCorrecton}
\end{equation}
where  our newly-introduced  tensor transition polarizability $\gamma$ depends on both the nuclear and  electronic structure. We have introduced an auxiliary rank-2 tensor $\{I\otimes{I}\}^{(2)}$ in front of the tensor polarizability to factor out the dependence on magnetic quantum numbers. Combined with this tensor term, 
Eq.~(\ref{eq:std_form}) is the most general parametrization of the transition polarizability as long as we only keep interactions linear in the static and laser fields. It is worth noting that in the second order, due to a particular selection of prefactors in Eq.~\eqref{eq:std_form}, $\alpha$ and $\beta$ do not depend on the hyperfine components of the initial and final states. We will demonstrate that the HFI-mediated corrections  would introduce the $F_i-$ and $F_f-$dependence to the scalar and vector polarizabilities. 

Based on these arguments, and taking into account the fact that the HFI is a scalar (see the discussion in Sec.~\ref{Sec:Third-order-HPI-trans-pol}),
we rewrite Eq.~\eqref{eq:std_form} in the following generalized form that now includes the tensor contribution~\eqref{Eq:GammaCorrecton}, as well as the $F$-dependence of the scalar and vector polarizabilities
\begin{align} \label{Eq:Tran-pol-tensorial}
    a_{i\rightarrow{f}}&=-\sqrt{3(2F_f+1)}w_0( \hat{\v{\varepsilon}},\hat{\v{e}})\alpha^{F_i \rightarrow F_f} \delta_{F_iF_f}\delta_{M_iM_f}\nonumber\\
    &-\sqrt{2}\brad{f}\sigma\ketd{i}w_1( \hat{\v{\varepsilon}},\hat{\v{e}}) \beta^{F_i \rightarrow F_f}
\nonumber\\
&+\brad{f}\{I\otimes{I}\}^{(2)}\ketd{i}w_2( \hat{\v{\varepsilon}},\hat{\v{e}})\gamma^{F_i \rightarrow F_f}
\,,
\end{align}
where we have used the Wigner-Eckart theorem and introduced the multipolar polarization weights~\cite{TraXiaDer23-2ndOrderBeta}
\begin{align}\label{eq:w_q}
w_Q( \hat{\v{\varepsilon}},\hat{\v{e}})&=(-1)^{Q}\sum\limits_{M_Q}(-1)^{M_Q+F_f-M_f}\nonumber\\
&\times\threej{F_f}{Q}{F_i}{-M_f}{-M_Q}{M_i}(\hat{\v{\varepsilon}}\otimes{\hat{\v{e}}})_{M_Q}^{(Q)}\,,
\end{align}
with $M_f$, ${M_Q}$, and $M_i$ being the magnetic quantum numbers. Note that selection rules fix the value of $M_Q = M_i -M_f$. 
The compound tensors of rank $Q$ for the two vectors $\hat{\v{\varepsilon}}$ and $\hat{\v{e}}$ are understood as
$\left(\hat{\v{\varepsilon}}\otimes\hat{\v{e}}\right)^{(Q)}_{M_Q}=\sum\limits_{\mu{\nu}}C_{1\mu1\nu}^{Q\,M_Q}\hat{\epsilon}_{\mu}\hat{e}_{\nu}$, where $C_{1\mu1\nu}^{Q\,M_Q}$ are Clebsch-Gordan coefficients and the $A_\mu$ components of a vector $\v{A}$  in the spherical (or helicity) basis expressed in terms of its Cartesian components as~\cite{VarMosKhe88}
$A_0 = A_z$, $A_{+1} = -  \left(A_x + i A_y \right)/\sqrt{2}$,  $A_{-1} =  \left(A_x - i A_y \right)/\sqrt{2}
$. In particular, the combinations of polarization vectors are $(\hat{\v{\varepsilon}}\otimes{\hat{\v{e}}})^{(0)} =- 
(\hat{\v{\varepsilon}}\cdot {\hat{\v{e}}})/\sqrt{3}$ and $(\hat{\v{\varepsilon}}\otimes{\hat{\v{e}}})^{(1)} =i 
(\hat{\v{\varepsilon}}\times {\hat{\v{e}}})/\sqrt{2}$, in agreement with Eq.~\eqref{eq:std_form}. We will consider the relevant components of the rank-2 tensor $(\hat{\v{\varepsilon}}\otimes{\hat{\v{e}}})^{(2)}$ in Sec.~\ref{Sec:ReinterpretationPNC}.

Here, we note that the reduced matrix elements of the auxiliary rank-2 tensor $\{I\otimes{I}\}^{(2)}$ present in Eq.~\eqref{Eq:Tran-pol-tensorial} is given by
\begin{align}
\brad{f}\{I\otimes{I}\}^{(2)}\ketd{i}&=(-1)^{2F_i-F_f+I-J_f}\sqrt{5}\,{[F_f,F_i]}^{1/2} \label{eq:fIIi}\nonumber\\
& \times I(I+1)[I]\sixj{1}{1}{2}{I}{I}{I} \delta_{J_iJ_f}\,,
\end{align}
where $[J_1,\,J_2,\,...\,J_n]\equiv(2J_1+1)(2J_2+1)\ldots(2J_n+1)$.
For our target $6S_{1/2} \rightarrow 7S_{1/2}$ transition in $\mathrm{^{133}Cs}$, the above expression evaluates to
\begin{equation}
    \brad{F_f}\{I\otimes{I}\}^{(2)}\ketd{F_i} =(-1)^{F_f}6\sqrt{35}[F_f,F_i]^{1/2}\,,
\end{equation}
which, for the special case where $F_{f,i}=3,4$, gives
\begin{subequations}
\label{Eq:II2numericalVals}
   \begin{align}
    \brad{3}\{I\otimes{I}\}^{(2)}\ketd{3} & = -42 \sqrt{35} \,, \\
    \brad{4}\{I\otimes{I}\}^{(2)}\ketd{4} & =  54 \sqrt{35} \,,\\
    \brad{3}\{I\otimes{I}\}^{(2)}\ketd{4} & = -126 \sqrt{5}  \,,\\
    \brad{4}\{I\otimes{I}\}^{(2)}\ketd{3} & =  126 \sqrt{5} \,.
\end{align} 
\end{subequations}
We will need these values in our analysis for $^{133}$Cs

To reiterate, due to the HFI, the scalar, vector, and tensor transition polarizabilities entering Eq.~\eqref{Eq:Tran-pol-tensorial} have an $F$-dependence of the form ($X=\alpha,\beta,\gamma$)
\begin{align}
    X^{F_i \rightarrow F_f} = X^{[2]} + \delta X^{F_i \rightarrow F_f}\,,
\end{align} 
where the second-order term $X^{[2]}$ is $F$-independent. For the $S_{1/2}  \rightarrow S_{1/2}$ transitions,  $\gamma^{[2]} \equiv 0$, thereby 
$\gamma^{F_i \rightarrow F_f} =\delta \gamma^{F_i \rightarrow F_f}$.  Expressions for the second-order scalar and vector transition polarizabilities, $\alpha^{[2]}$ and $\beta^{[2]}$, can be found, e.g., in Ref.~\cite{TraXiaDer23-2ndOrderBeta}. Substantial attention~\cite{Dzuba1997,Safronova1999,Vas2002,Dzuba2002,Toh2019,Ben1999,DzuFla00,sahoo20,TraXiaDer23-2ndOrderBeta} has been paid over the years to determining their accurate values since they are required for interpreting the results of APV experiments.  As the reference values for the second-order polarizabilities for the $6S_{1/2} \rightarrow 7S_{1/2}$ transition in $^{133}$Cs, we use values computed recently by our group~\cite{TraXiaDer23-2ndOrderBeta}
 \begin{subequations}
   \label{Eq:RecVals}
\begin{align}
\alpha^{[2]}&= -266.31(23)\,, \\
\beta^{[2]}  &= 26.912(30) \,, \\
\gamma^{[2]}  &= 0 \,.         
\end{align}
\end{subequations}
These values are in atomic units, $a_0^3$, where $a_0$ is the Bohr radius. 

We now proceed to the derivation of the hyperfine corrections $\delta \alpha^{F_i \rightarrow F_f}$, $\delta \beta^{F_i \rightarrow F_f}$, and $\delta \gamma^{F_i \rightarrow F_f}$ to  transition polarizabilities.


\section{Hyperfine corrections to transition polarizabilities} \label{Sec:Third-order-HPI-trans-pol}

To evaluate the hyperfine-mediated corrections to the transition polarizability, we follow the third-order formalism developed in Refs.~\cite{BelSafDer06,DzuFlaBel10,RosGheDzu09}. References~\cite{BelSafDer06,DzuFlaBel10} computed the static differential polarizabilities for transitions between levels of the hyperfine manifold attached to the $S_{1/2}$  ground state. Reference~\cite{RosGheDzu09}
generalized that formalism to the evaluation of dynamic (AC) polarizabilities. 
These papers  focused on the characterization of clock shifts, which formally map into the evaluation of the diagonal matrix elements of the transition amplitude. Here we further generalize our earlier formalism and consider {\em off-diagonal}  matrix elements of the transition amplitude. In the context of APV, Ref.~\cite{DeMille1998} has considered transition polarizabilities (including tensor contribution) for transitions between hyperfine components attached to the Cs ground state.

\begin{figure}[!ht]
    \centering
    \includegraphics[width=0.75\columnwidth]{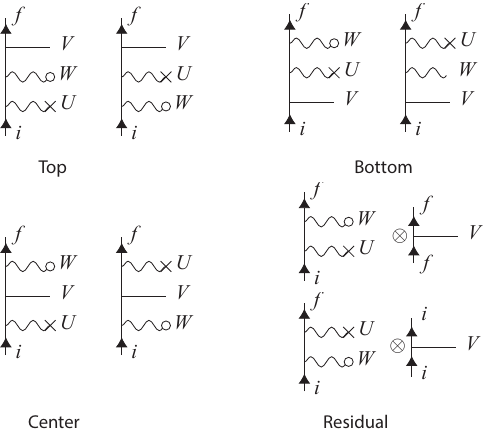}
    \caption{Top, Bottom, Center, and Residual
(Normalization) diagrams contributing to the
hyperfine-mediated corrections to transition polarizability. Here, $i$ and $f$ denote the initial and
the final states; $W$, $U$, and $V$ represent the electric dipole  $-\hat{\v{\varepsilon}}\cdot{\v{D}}$, $-\hat{\v{e}}\cdot{\v{D}}$, and the hyperfine interaction $V^{\rm{HFI}}$, respectively.}
    \label{fig:TCBN}
\end{figure}

The four relevant diagrams representing third-order contributions to the $i\rightarrow f$ transition amplitude, top (T), center (C), bottom (B), and residual (R), are shown in Fig.~{\ref{fig:TCBN}}, with each diagram involving one hyperfine interaction and two $E1$ interactions (one with the laser, and another one with the static field). These diagrams are named after the position of the hyperfine interaction in the string of three operators. Explicitly, these terms read
\begin{subequations}
    \begin{align}\label{eq:TBCN_operators}
T_{i\rightarrow{f}}&=\sum\limits_{ab}\frac{V_{fa}^{\rm{HFI}}(\hat{\v\varepsilon}\cdot\v{D}_{ab})(\hat{\v{e}}\cdot\v{D}_{bi})}{\Delta{E_{fa}}\Delta{E_{ib}}}\nonumber\\
&+\sum_{ab}\frac{V_{fa}^{\rm{HFI}}(\hat{\v{e}}\cdot{\v{D}_{ab}})(\hat{\v{\varepsilon}}\cdot{\v{D}_{bi}})}{\Delta{E_{fa}}\Delta{E_{fb}}}\,,\\
B_{i\rightarrow{f}}&=\sum\limits_{ab}\frac{(\hat{\v{\varepsilon}}\cdot{\v{D}_{fa}})(\hat{\v{e}}\cdot{\v{D}_{ab}})V_{bi}^{\rm{HFI}}}{\Delta{E_{ia}}\Delta{E_{ib}}}\nonumber\\
&+\sum_{ab}\frac{(\hat{\v{e}}\cdot{\v{D}_{fa}})(\hat{\v{\varepsilon}}\cdot{\v{D}_{ab}})V_{bi}^{\rm{HFI}}}{\Delta{E_{fa}}\Delta{E_{ib}}}\,,\\
C_{i\rightarrow{f}}&=\sum\limits_{ab}\frac{(\hat{\v{\varepsilon}}\cdot{\v{D}_{fa}})V_{ab}^{\rm{HFI}}(\hat{\v{e}}\cdot{\v{D}_{bi}})}{\Delta{E_{ia}}\Delta{E_{ib}}}\nonumber\\
&+\sum_{ab}\frac{(\hat{\v{e}}\cdot{\v{D}_{fa}})V_{ab}^{\rm{HFI}}(\hat{\v{\varepsilon}}\cdot{\v{D}_{bi}})}{\Delta{E_{fa}}\Delta{E_{fb}}}\,,\\
R_{i\rightarrow{f}}&=-V_{ii}^{\rm{HFI}}\sum\limits_{a}\frac{(\hat{\v{\varepsilon}}\cdot{\v{D}_{fa}})(\hat{\v{e}}\cdot{\v{D}_{ai}})}{(\Delta{E_{ia}})^2}\nonumber\\
&-V_{ff}^{\rm{HFI}}\sum\limits_{a}\frac{(\hat{\v{e}}\cdot{\v{D}_{fa}})(\hat{\v{\varepsilon}}\cdot{\v{D}_{ai}})}{(\Delta{E_{fa}})^2}\,,
\end{align}
\end{subequations}
where $\Delta E_{ij}\equiv E_i-E_j$.
Note that the two terms inside each combination differ by the swap of the two polarization vectors $\hat{\v{\varepsilon}}$ and $\hat{\v{e}}$. Otherwise, the structure of the terms is similar. Further, the bottom and top diagrams are related as $B_{f\leftrightarrow{i}}=T^*_{i\leftrightarrow{f}}$.

Before carrying out the angular reduction of the expressions above , we briefly review the hyperfine interaction present in Eqs.~\eqref{eq:TBCN_operators}.
Following notation of Ref.~\cite{Joh07},
the  interaction of electrons with nuclear multipolar moments may be expressed as
 \begin{equation} 
 V^\mathrm{HFI}=\sum\limits_{N}\mathcal{T}^{(N)}\cdot\mathcal{N}^{(N)}\,,\ \label{Eq:HF-general}
 \end{equation}
where the rank-$N$ tensors $\mathcal{T}^{(N)}$ act in the electron space, and $\mathcal{N}^{(N)}$ act in the nuclear space. Note that $V^\mathrm{HFI}$ is a scalar ITO.
The nuclear reduced matrix elements 
$\brad{\gamma{I}}\mathcal{N}^{(N)}\ketd{\gamma{I}}$ are expressed in terms of the conventional nuclear magnetic-dipole ($M1$)  $\mu$ and  electric-quadrupole ($E2$) $Q$ moments  as
\begin{subequations}
    \begin{align}
\brad{\gamma{I}}\mathcal{N}^{(1)}\ketd{\gamma{I}} &= \sqrt{\frac{(2I+1)(I+1)}{I}}\mu \,,\\
\brad{\gamma{I}}\mathcal{N}^{(2)}\ketd{\gamma{I}} &=  
\sqrt{\frac{(2I +1)(I+1)(2I+3)}{ 4I(2I -1)}} Q \,.
\end{align}
\end{subequations}

Here the magnetic-dipole moment $\mu\equiv g_I I\mu_N$ with $\mu_N$ being the nuclear magneton and $g_I$ being the gyromagnetic ratio. For $^{133}$Cs, $g_I=0.73714$.
As for the nuclear electric-quadrupole moment $Q$, the measured hyperfine constant $B$ can be used to extract its value using theoretical values of the hyperfine electronic matrix elements. However, different measurements of $B$  yield different determinations. For instance, the measured~\cite{GerDerTan03} hyperfine 
constant $B$ in the $^{133}$Cs $6P_{3/2}$ state is $-0.4934{(17)}\, \mathrm{MHz}$, which differs from a more recent result~\cite{DasNat08} of, $-0.5266{(57)} \, \mathrm{MHz}$ by about $7\%$. 
Because the uncertainty in $B$ of Ref.~{\cite{GerDerTan03}} is smaller, we simply adopt the value $Q=-3.55(4)\mathrm{\,mbarn}$ therefrom.
Moreover, we find that the nuclear quadrupole contributions to the transition polarizabilities are suppressed compared to those due to the magnetic-dipole hyperfine interaction. 
For the same reason, we neglect even higher-rank nuclear multipoles, such as the poorly known magnetic octupole moment~\cite{GerDerTan03}, due to their diminishing role as compared to the magnetic-dipole contribution. 

 To flesh out the tensorial structure of the transition polarizability resulting from the diagrams~\eqref{eq:TBCN_operators}, we use the same re-coupling angular momentum algebra technique as in our derivation of the second-order expressions~\cite{TraXiaDer23-2ndOrderBeta}. 
 Since the HFI is a scalar ITO, the resulting tensorial structure of the transition polarizability is indeed given by Eq.~\eqref{Eq:Tran-pol-tensorial}.
The hyperfine corrections to transition polarizabilities are therefore given by
\begin{subequations}\label{Eq:HFI-corrections-polariz}
\begin{align}
\delta{\alpha}^{F_i \rightarrow F_f}&=-\frac{\brad{f}T^{(0)}+B^{(0)}+C^{(0)}+R^{(0)}\ketd{i}}{\sqrt{3(2F_f+1)}}\,,\\
\delta{\beta}^{F_i \rightarrow F_f}&=-\frac{\brad{f}T^{(1)}+B^{(1)}+C^{(1)}+R^{(1)}\ketd{i}}{\sqrt{2}\brad{f}\sigma\ketd{i}}\,,\\
\delta{\gamma}^{F_i \rightarrow F_f}&=\frac{\brad{f}T^{(2)}+B^{(2)}+C^{(2)}+R^{(2)}\ketd{i}}{\brad{f}\{I\otimes{I}\}^{(2)}\ketd{i}}\,.
\end{align}
\end{subequations}

We remind the reader that the various transition polarizabilities entering Eq.~\eqref{Eq:Tran-pol-tensorial} are assembled as 
$ 
X^{F_i \rightarrow F_f} = X^{[2]} + \delta X^{F_i \rightarrow F_f}\,,
$ 
where the second-order term $X^{[2]}$ is $F$-independent. We listed our recommended values~\cite{TraXiaDer23-2ndOrderBeta} for the second-order transition polarizabilities in Eqs.~\eqref{Eq:RecVals}.

 
The reduced matrix elements of individual diagrams entering Eqs.~\eqref{Eq:HFI-corrections-polariz} are given by
\begin{subequations}
\begin{align}
\brad{f}T^{(Q)}\ketd{i}&=\sum\limits_{NJ_aJ_b}(-1)^{F_f-F_i+J_a+J_i}[F_f,F_i,Q]^{1/2}\nonumber\\
&\times\sixj{F_f}{I}{J_a}{N}{J_f}{I}\sixj{Q}{J_i}{J_a}{I}{F_f}{F_i}\sixj{Q}{J_i}{J_a}{J_b}{1}{1}\nonumber\\
&\times\left\{S_T^{(J_aJ_bN)}[fi]+(-1)^QS^{(J_aJ_bN)}_T[ff]\right\}\,,\label{eq:top}\\
\brad{f}B^{(Q)}\ketd{i}&=\sum\limits_{NJ_aJ_b}(-1)^{J_i+J_b}[F_f,F_i,Q]^{1/2}\nonumber\\
&\times\sixj{F_i}{I}{J_b}{N}{J_i}{I}\sixj{Q}{J_f}{J_b}{I}{F_i}{F_f}\sixj{Q}{J_b}{J_f}{J_a}{1}{1}\nonumber\\
&\times\left\{S_B^{(J_aJ_bN)}[ii]+(-1)^QS_B^{(J_aJ_bN)}[fi]\right\}\,,\label{eq:bottom}\\
\brad{f}C^{(Q)}\ketd{i}&=
\sum\limits_{NJ_aJ_b}(-1)^{J_a-J_i+F_i-F_f+N+1} [F_f,F_i,Q]^{1/2} \nonumber\\
&\times\sum_j [j] \ninej{J_f}{J_i}{j}{F_f}{F_i}{Q}{I}{I}{N}\ninej{J_f}{J_i}{j}{1}{1}{Q}{J_a}{J_b}{N}\nonumber\\
&\times\left\{S_C^{(J_aJ_bN)}[ii]+(-1)^{Q}S_C^{(J_aJ_bN)}[ff]\right\}\,,\label{eq:center}\\
\brad{f}R^{(Q)}\ketd{i}&=(-1)^{2F_f-I+F_i+J_i+1}[F_f,F_i,Q]^{1/2}\nonumber\\
\label{eq:norm}
&\times\sixj{Q}{J_f}{J_i}{I}{F_i}{F_f}\sum\limits_{J_a}\sixj{Q}{J_i}{J_f}{J_a}{1}{1}\nonumber\\
&\times\left\{V[i]S_R^{J_a}[f]+(-1)^{Q}V[f]S_R^{J_a}[i]\right\}\,,
\end{align}
\end{subequations}
which are expressed in terms of the reduced sums
\begin{widetext}
\begin{subequations}
\begin{align}
    S_T^{(J_aJ_bN)}[\alpha\beta] & ={\sum\limits_{n_{a}n_{b}}\frac{\brad{I}\mathcal{N}^{(N)}\ketd{I}\brad{n_fJ_f}\mathcal{T}^{(N)}\ketd{n_{a}J_a}\brad{n_aJ_a}D\ketd{n_bJ_b}\brad{n_bJ_b}D\ketd{n_iJ_i}}{\Delta{E_{\alpha{a}}}\Delta{E_{\beta{b}}}}}\,,\\
S_B^{(J_aJ_bN)}[\alpha\beta] & =\sum\limits_{n_an_b}\frac{\brad{n_fJ_f}D\ketd{n_aJ_a}\brad{n_aJ_a}D\ketd{n_bJ_b}\brad{I}\mathcal{N}^{(N)}\ketd{I}\brad{n_bJ_b}\mathcal{T}^{(N)}\ketd{n_iJ_i}}{\Delta{E_{\alpha{a}}}\Delta{E_{\beta{b}}}}\,,\\
S_C^{(J_aJ_bN)}[\alpha\beta] & =\sum\limits_{n_{a}n_{b}}\frac{\brad{n_{f}J_f}D\ketd{n_aJ_a}\brad{I}\mathcal{N}^{(N)}\ketd{I}\brad{n_aJ_a}\mathcal{T}^{(N)}\ketd{n_bJ_b}\brad{n_bJ_b}D\ketd{n_iJ_i}}{\Delta{E_{\alpha{a}}}\Delta{E_{\beta{b}}}}\,,\\
S_{R}^{(J_a)}[\alpha] &= \sum_{n_a} \frac{\brad{n_{f}J_{f}}D\ketd{n_aJ_a}\brad{n_aJ_a}D\ketd{n_iJ_i}}{(\Delta E_{\alpha a})^2}\,,
\end{align}
\end{subequations}
\end{widetext}
and the HFI diagonal matrix elements
\begin{align}
V[\alpha]=&(-1)^{I+J_{\alpha}+F_{\alpha}}\sum\limits_{N}\sixj{F_{\alpha}}{J_{\alpha}}{I}{N}{I}{J_{\alpha}}\nonumber\\
&\times\brad{n_{\alpha}\,J_{\alpha}}\mathcal{T}^{(N)}\ketd{n_{\alpha}\,J_{\alpha}}\brad{I}\mathcal{N}^{(N)}\ketd{I}\,.    
\end{align}

\section{Numerical results for hyperfine corrections}\label{Sec:Numerical}

\begin{table}[ht!]
        \centering
    \begin{tabular}{ccrr} 
        \hline\hline
    \multicolumn{1}{c}{Transition} &
    \multicolumn{1}{c}{}&
    RPA(BO) & Semi-emp.\\   
    \hline
  &$\delta \alpha$   & $-6.779[-3]$ & $-7.315[-3]$\\
$6S_{1/2}\,F=3\rightarrow{7S_{1/2}\,F=3}$  & 
$\delta\beta$  & $2.681[-3]$&$2.735[-3]$\\
  & $\delta\gamma$& $-1.209[-5]$ &  $-1.145[-5]$\\
  \hline
  & $\delta\alpha$ & $0$& $0$\\
$6S_{1/2}\,F=4\rightarrow7S_{1/2}\,F=3$  &$\delta\beta$ &$-1.978[-4]$ &   {$-1.272[-4]$}\\
  &$\delta\gamma$  & $-1.849[-5]$& {$-1.748[-5]$}\\
  \hline
  & $\delta\alpha$ & $0$& $0$\\
 $6S_{1/2}\,F=3\rightarrow7S_{1/2}\,F=4$ & $\delta\beta$ &$7.937[-4]$&  {$7.351[-4]$}\\
  & $\delta\gamma$  & $-1.849[-5]$&{$-1.748[-5]$}\\
  \hline
  & $\delta\alpha$  
   &$5.273[-3]$&  { $5.689[-3]$}\\
 $6S_{1/2}\,F=4\rightarrow7S_{1/2}\,F=4$ &$\delta\beta$ 
  & $-2.086[-3]$
 &    $-2.127[-3]$ \\
  & $\delta\gamma$
  &$-1.804[-5]$ & {$-1.706[-5]$} \\
  \hline\hline
    \end{tabular}
    \caption{
   Hyperfine  corrections to the scalar, $\delta{\alpha}^{F_i \rightarrow F_f}$, vector, $\delta{\beta}^{F_i \rightarrow F_f}$,  and tensor, $\delta{\gamma}^{F_i \rightarrow F_f} ={\gamma}^{F_i \rightarrow F_f} $, transition polarizabilities for the indicated transitions in $^{133}\mathrm{Cs}$.
   Semi-empirical values are anticipated to have better accuracy than the RPA(BO) results. 
    The notation $x[y]$ stands for $x\times 10^y$.}
    \label{Tab:Third-order-polarizabilities} 
\end{table}

In Sec.~\ref{Sec:Third-order-HPI-trans-pol}, we have presented the formulation for the hyperfine corrections to the scalar, vector and tensor transition polarizabilities. In this section, we our numerical results, which are compiled in Table~\ref{Tab:Third-order-polarizabilities}.
To arrive at these values, we employed relativistic many-body methods for computing atomic structure. A detailed discussion of these methods and their numerical implementation can be found in Refs.~\cite{TraXiaDer23-2ndOrderBeta,tan2023precision} and references therein. 
Simply put, we used the frozen-core $V^{N-1}$ Dirac-Hartree-Fock (DHF), Brueckner orbitals (BO), and random phase approximation (RPA).  Among these approximations the RPA(BO) approach is the most complete as it incorporates  the core polarization and the core  screening effects. The RPA(BO)  results are listed in Table~\ref{Tab:Third-order-polarizabilities}. In our calculations, we use a dual-kinetic balance B-spline DHF basis set~\cite{BelDer08.DKB} containing $N=60$ basis functions of order $k=9$ per partial wave generated in a cavity of radius $R_{\mathrm{max}}=250\,\mathrm{a.u.}$,  the same as in Refs.~\cite{TraXiaDer23-2ndOrderBeta,tan2023precision}. 

To improve the accuracy of our calculations, we also employ a semi-empirical approach. To this end, we point out that there are three  atomic properties entering the reduced sums: the energies, the $E1$ matrix elements, and the HFI matrix elements. 
We therefore replace a certain subset of {\em ab initio} RPA(BO) quantities with the experimental or other high-accuracy values. Determining this subset, however, requires some care. Indeed, although the low-$n$ orbitals from the finite basis set closely resemble those obtained with the conventional finite-difference technique computed with practically infinite cavity radius, as $n$ increases, the mapping of the basis states to physical states deteriorates. In our basis set, we find the boundary for the transition from physical to non-physical orbitals to be at the radial quantum number $n_r=12$, without loss of numerical accuracy for matrix elements and energies. Because of this, while evaluating the reduced sums, we use 
 the NIST recommended~\cite{NIST_ASD}  energies for the physical states, $n_{a,\,b}P_J$ with $n_{a,\,b}=6-12$ and $n_{a,\,b}D_J$ with $n_{a,\,b}=5-11$. For the same reasons, we replace the RPA(BO) $E1$  matrix elements for the $6S_{1/2}\rightarrow n_{a,\,b}P_J$ and  $7S_{1/2}\rightarrow n_{a,\,b}P_J$ channels  with their experimental values tabulated in Ref.~\cite{Toh2019} for $n_{a,\,b}=6,\,7$ and with their high-accuracy relativistic coupled-cluster counterparts~\cite{tan2023precision} for $n_{a,\,b}=8-12$.



The semi-empirical matrix elements of the hyperfine interaction involve  ``physical'' states with principle quantum numbers $6\le{n_{a,\,b}}\le{12}$. We evaluate them as follows.
The diagonal hyperfine matrix elements are extracted from the experimental values~\cite{Belin1976,AriIngVio77,Auzinsh2007} of hyperfine constants $A$ from the relation,
\begin{align}
    A=\langle{\mathcal{T}^{(1)}}\rangle_J\langle\mathcal{N}^{(1)}\rangle_I/(IJ)\,,
\end{align}
where $\langle{\mathcal{T}^{(1)}}\rangle_J$ and $\langle\mathcal{N}^{(1)}\rangle_I$
are the so-called stretched matrix elements expressed in terms of the reduced matrix elements
\begin{equation}
\langle{\mathcal{O}^{(N)}}\rangle_J=\begin{pmatrix}
    J & N & J\\
    -J & 0 & J
\end{pmatrix}\brad{\gamma J}\mathcal{O}^{(N)}\ketd{\gamma J}\,.
\end{equation}
The off-diagonal HFI matrix elements between the $S_{1/2}$ states 
were evaluated as the geometric mean of the diagonal matrix elements~\cite{DzuFla00}
\begin{align}
\bra{n'S_{1/2}}V^{\rm{HFI}}\ket{nS_{1/2}}&=\bra{n'S_{1/2}}V^{\rm{HFI}}\ket{n'S_{1/2}}^{1/2}\nonumber\\
&\times\bra{nS_{1/2}}V^{\rm{HFI}}\ket{nS_{1/2}}^{1/2}\,,
\end{align}
where the diagonal matrix elements come from the experimental values of the hyperfine constant $A$. The high accuracy of this approximation has been confirmed in Ref.~\cite{Derevianko1999b}. The remaining off-diagonal magnetic-dipole HFS matrix elements between the ``physical'' states were determined using the relativistic coupled-cluster method, with the code described in~\cite{tan2023precision}. As for the nuclear quadrupole HFI contributions,  we found them to be suppressed.  Thereby, we kept their RPA(BO) matrix elements.


Our numerical results for the hyperfine corrections to transition polarizabilities are listed in Table~\ref{Tab:Third-order-polarizabilities}.
Overall, the corrections to the  polarizabilities are below the $10^{-2}\, \mathrm{a.u.}$ level. 
The $\delta \alpha$ corrections are identically zero for the $F_i \neq F_f$ transitions  due to the scalar nature of the underlying ITO. Otherwise, 
$|\delta \alpha^{F_i\rightarrow F_f}| \sim 5 \times 10^{-3}$ is about 5 orders of magnitude smaller than  $|\alpha^{[2]}| \approx 3 \times 10^2$. As to the vector transition polarizability, $|\delta \beta^{F_i\rightarrow F_f}|$ is about 4-5 orders of magnitude smaller than $|\beta^{[2]}| \approx 3 \times 10^1$. The $|\delta \beta^{F_i\rightarrow F_f}|$ corrections to the $F_i=F_f$ transitions are an order of magnitude larger than those for the  $F_i\neq F_f$ transitions. We observe that the tensor transition polarizability, $\gamma^{F_i \rightarrow F_f} =\delta \gamma^{F_i \rightarrow F_f}$, is in the order of ${10}^{-5}\,\mathrm{a.u.}$ The relative smallness of the numerical values of the tensor transition polarizabilities as compared to their scalar and vector counterparts is due, in part, to the large values, $\sim 3\times 10^2$, of the prefactors
$\brad{F_f}\{I\otimes{I}\}^{(2)}\ketd{F_i}$ in Eq.~\eqref{Eq:II2numericalVals}.
Further, $\gamma^{3 \rightarrow 4} = \gamma^{4 \rightarrow 3}$ as can be proven by a direct examination of our analytical expressions.
Finally, the difference between our RPA(BO) and  semi-empirical estimates does not exceed 10\%, which we take as the uncertainty of our results.

\section{Discussion}\label{Sec:Res-and-Dis}

We have presented the theoretical formulation and numerical estimate for the hyperfine corrections to the transition polarizabilities. In this section, we investigate the impact of neglecting the hyperfine-mediated tensor polarizability $\gamma$ and the hyperfine-state dependence of the scalar $\alpha$, and vector $\beta$ polarizabilities on the extraction of electroweak observables from  APV experiments. In particular, we reanalyze two Boulder experiments~\cite{WooBenCho97,ChoWooBen97} and
compute corrections to their extracted value of the $^{133}\mathrm{Cs}$ anapole moment and the ratio $\alpha/\beta$ of the scalar and vector transition polarizabilities.

\subsection{Reinterpretation of the Boulder parity violation measurement}
\label{Sec:ReinterpretationPNC}
We start by reviewing the Boulder APV experiment~\cite{WooBenCho97,WooBenRob99} and the assumptions that went into its analysis.
The experiment utilized the Stark interference technique to extract the ratio of the PNC amplitude to the vector transition polarizability, $\mathrm{Im(E1_{PNC})}/\beta$. Notice the use of $\beta$ without specifying hyperfine components, as the hyperfine corrections were neglected. It is our goal to introduce $F$-dependent corrections to $\beta$ here.

The Boulder experiment used a spin-polarized $^{133}$Cs beam subjected to a uniform and static  electric field, with a laser driving the nominally $E1$-forbidden transition between various hyperfine components of the ground $6S_{1/2}$ and the excited  $7S_{1/2}$ state. The DC electric field opens up an $E1$ transition channel between these states by mixing the $S$ and $P$ states. The total transition rate $R$ is determined by a combination of the Stark-induced, parity-violating (PNC), and $M1$ transition amplitudes
\begin{equation}
 R={|A^{\rm{Stark}}_{i\rightarrow{f}}+A^{\rm{PNC}}_{i\rightarrow{f}}+A^{\rm{M_1}}_{i\rightarrow{f}}|^2} \,, 
\label{Eq:TheRate}  
\end{equation}
where~\cite{WooBenRob99}
\begin{subequations}\label{eq:amplitudes}
    \begin{align}
A^{\mathrm{Stark}}_{i\rightarrow{f}}&=\alpha{\v{\mathbfcal{E}}_L\cdot{\v{\mathbfcal{E}}_S}}\,\delta_{F_fF_i}\delta_{M_fM_i}\nonumber\\
&+i\beta({\v{\mathbfcal{E}}_S\times{\v{\mathbfcal{E}}_L})\cdot{\bra{f}\v{\sigma}\ket{i}}},\label{eq:stark}\\
A^{\mathrm{PNC}}_{i\rightarrow{f}}&=i\mathrm{Im}(E1_{\rm PNC}){\v{\mathcal{E}}_S}\cdot{\bra{f}\v{\sigma}\ket{i}},\label{eq:APNC}\\
A^{M1}_{i\rightarrow{f}}&=(M1)_\mathrm{rad} (\hat{\v{k}}_L\times{{\v{\mathcal{E}}}_S})\cdot{\bra{f}\v{\sigma}\ket{i}}\,.\label{eq:AM1}
    \end{align}
\end{subequations}
Here we have changed the notation of Ref.~\cite{WooBenRob99} to be consistent with that of the previous sections. 

In Eqs.~\eqref{eq:amplitudes}, $\v{\mathbfcal{E}}_L = \mathcal{E}_L\hat{\v{\varepsilon}} $ is the laser field driving the transition with $\hat{\v{k}}_L$ being a unit vector in its propagation direction, $\v{\mathcal{E}}_S=\mathcal{E}_S\hat{\v{e}}$ is the DC electric field, and $\alpha$ and $\beta$ are the scalar and vector transition polarizabilities, introduced in earlier sections. To set the stage, for now, as in Ref.~\cite{WooBenRob99}, we neglected the $F$-dependence of $\alpha$ and $\beta$ and omitted the tensor $(\gamma)$ contribution.  The PNC amplitude $E1_{\mathrm{PNC}}$ includes both the nuclear spin-dependent and spin-independent effects and $(M1)_\mathrm{rad}$ stands for the radial integral of the $6S_{1/2}-7S_{1/2}$ $M1$ matrix element~\cite{WooBenRob99}. We will neglect the $A_{i\rightarrow{f}}^{M1}$ $M1$ amplitude for reasons discussed in Ref.~{\cite{WooBenRob99}}.

The Stark interference technique amplifies the feeble PNC amplitude $A^{\mathrm{PNC}}_{i\rightarrow{f}}$ with the help of the much stronger $A^{\mathrm{Stark}}_{i\rightarrow{f}}$ amplitude: the interference between $A^{\mathrm{PNC}}_{i\rightarrow{f}}$ and $A^{\mathrm{Stark}}_{i\rightarrow{f}}$ manifests itself as a cross term when expanding the square in the rate expression, Eq.~\eqref{Eq:TheRate}. To access this Stark-PNC interference term, 
the experiment~\cite{WooBenRob99} involved measuring the change in the transition rate $R$, Eq.~\eqref{Eq:TheRate}, under various parity reversals, which included flipping
 the direction of the applied DC electric field, flipping the sign of the relevant component of the laser polarization, or changing the sign of the magnetic quantum numbers~\cite{WooBenRob99}. The PNC amplitude was extracted from two transition rates, $R^+$ and $R^-$ measured under opposite parities.  
 A parity reversal results in a sign flip of the PNC amplitude $A^{\rm{PNC}}_{i\rightarrow{f}}$, while leaving the sign of the Stark-induced amplitude $A^{\rm{Stark}}_{i\rightarrow{f}}$ unaffected. 
 
 The Stark-induced amplitude $A^{\mathrm{Stark}}_{i\rightarrow{f}}$ in Eq.~\eqref{eq:stark}  generally depends on both the scalar and vector polarizabilities. However, in the Boulder experiment, the transitions were driven between the states of different values of $F$ ($F_i\neq F_f$), and thereby, only the vector polarizability contribution remained in Eq.~\eqref{eq:stark}. Therefore, it is the vector polarizability $\beta$ that enters the interference term with $E1_{\rm{PNC}}$.
Explicitly, the PNC amplitude was extracted from the normalized difference in the two transition rates,
\begin{align}\label{eq:Ratio_diff_over_sum}
      \frac{R^+-R^-}{R^++R^-}\propto{\frac{\mathrm{Im(E1_{\rm{PNC})}}}{{\beta}}}\,.
 \end{align}

Next, we specify the geometry of the Boulder experiment~\cite{WooBenCho97,WooBenRob99}. In the setup of the Boulder experiment, a $^{133}$Cs atomic beam travels along the $z$-axis and an externally-applied magnetic field is aligned along the beam propagation direction, defining the quantization axis. Before entering the excitation-laser interaction region, the Cs atoms are optically pumped into the ``stretched'' hyperfine sub-levels of the $6S_{1/2}$ ground states, either $F_i=3,\,M_i=\pm{3}$ or  $F_i=4,\,M_f=\pm{4}$. The transitions to the $7S_{1/2}$ hyperfine manifold are driven by a standing wave laser with the cavity axis aligned along the $y$-axis. The excitation laser field $\mathbfcal{E}_L$ is elliptically polarized,  $\mathbfcal{E}_L=\mathcal{E}_L^z\hat{\v{z}}+i\mathcal{E}_L^I\hat{\v{x}}$. Finally, a static and uniform electric field $\v{\mathbfcal{E}}_S=\mathcal{E}_S^x\hat{\v{x}}$ is aligned along the $x$-axis. 

Having reviewed the Boulder experiment, now we examine the effect of our newly-introduced tensor transition polarizability $\gamma$, as well as the nuclear-spin-dependent corrections to $\alpha$ and $\beta$, and assess whether they affect the extraction of the PNC amplitude $E1_{\rm{PNC}}$. To this end, we rewrite Eq.~\eqref{Eq:Tran-pol-tensorial} as
\begin{align}\label{eq:revised_amplitude}
A_{i\rightarrow{f}}^{\mathrm{Stark}}&=
\alpha^{F_i\rightarrow F_f}{{\mathbfcal{E}_L}\cdot{{\bm{\mathbfcal{E}}_S}}}\delta_{F_fF_i}\delta_{M_fM_i}\nonumber\\
&+i\beta^{F_i\rightarrow F_f}(\bm{\mathbfcal{E}_L}\times{{\bm{\mathbfcal{E}_S}}})\cdot \bra{f}\v{\sigma}\ket{i}\nonumber\\
&+\gamma^{F_i\rightarrow F_f}w_2( \hat{\v{\varepsilon}},\hat{\v{e}}) \mathcal{E}_L\mathcal{E}_S
\brad{f}\{I\otimes{I}\}^{(2)}\ketd{i}\,, 
\end{align}
where we have again used $A_{i\rightarrow{f}}^{\mathrm{Stark}} = \mathcal{E}_L\mathcal{E}_S \, a_{i\rightarrow{f}} $.
The reduced matrix element $\brad{f}\{I\otimes{I}\}^{(2)}\ketd{i}$ is again given by Eq.~\eqref{eq:fIIi} and the polarization and state-dependent factor is, explicitly (c.f. Eq.~\eqref{eq:w_q})
\begin{align}
 w_2( \hat{\v{\varepsilon}},\hat{\v{e}})&=\sum\limits_{M_Q=-2}^{2}(-1)^{M_Q+F_f-M_f}  \label{Eq:w2}\\ 
&\times{\threej{F_f}{2}{F_i}{-M_f}{-M_Q}{M_i}}(\hat{\v{\varepsilon}}\otimes\hat{\v{e}})^{(2)}_{M_Q}\,.\nonumber   
\end{align}
The components of the rank-two compound tensor of  electric field polarizations are
\begin{equation}
  (\hat{\v{\varepsilon}}\otimes\hat{\v{e}})^{(2)}_{M_Q} = 
  \sum_{\mu,\lambda=-1}^1 C^{2M_Q}_{1\mu1\lambda}  
  \hat{\v{\varepsilon}}_\mu \hat{\v{e}}_\lambda \,.
\end{equation}

Note that the selection
rules for the $3j$-symbol fix $M_Q = M_i - M_f$ in Eq.~\eqref{Eq:w2}. Moreover, since we are interested in transitions between stretched hyperfine states $\ket{F,M_F=\pm F}$ with $F_i = F_f\pm1$, only terms with $M_Q = \pm 1$ survive in Eq.~\eqref{Eq:w2}.
For the Boulder experiment where $\hat{\v{\varepsilon}} = \varepsilon_L^z \hat{\v{z}} + i \varepsilon_L^I \hat{\v{x}} $ and
$\hat{\v{e}} = \hat{\v{x}}$, we find the needed components of the second-rank tensor to be $(\hat{\v{\varepsilon}}\otimes\hat{\v{e}})^{(2)}_{\pm 1} = \mp \frac{1}{2}\varepsilon_L^z$.
%
%
Then the Stark-induced  amplitude for transitions between stretched states with $F_i =F_f\pm1$ can be simplified to
\begin{align}\label{eq:revised-Stark-ind-amp}
A^{\mathrm{Stark}}_{i\rightarrow{f}}
&=\beta^{F_i \rightarrow F_f}{\mathcal{E}_L^z{\mathcal{E}_S^x}}C^{F_fM_f}_{F_iM_f\pm1}\nonumber\\
&\pm\gamma^{F_i \rightarrow F_f}\mathcal{E}^z_L{\mathcal{E}^x_S}U^{F_fM_f}_{F_iM_f\pm1}/2\,,
\end{align}
where $\mathcal{E}_L^z$ and $\mathcal{E}_S^x$ are the components of the laser and the applied DC electric fields, respectively. The coefficients $C^{F_fM_f}_{F_iM_f\pm1}$ are defined as
\begin{align}
    C^{F_fM_f}_{F_iM_f\pm1}&=\frac{(-1)^{I+S+F_i+1}}{2\sqrt{3}} [F_f,F_i]^{1/2}\nonumber\\
    &\times\sixj{1/2}{F_f}{I}{F_i}{1/2}{1} \threej{F_f}{1}{F_i}{-M_f}{\mp1}{M_f \pm1}\,,
\end{align}
and are tabulated in Ref.~\cite{Gilbert1986}. Here we introduced a similar coefficient,
\begin{align}
U^{F_fM_f}_{F_i M_f\pm1}&=(-1)^{F_f-M_f}\threej{F_f}{2}{F_i}{-M_f}{\mp{1}}{M_f\pm1}\nonumber\\ &\times\bra{F_f}|\{I\otimes{I}\}^{(2)}|\ket{F_i}\,,
\end{align}
which specifies the dependence of the tensor contribution on the magnetic quantum numbers. 
The ``$\pm$'' signs appearing in the  $C^{F_fM_f}_{F_iM_f\pm1}$ and $U^{F_f M_f}_{F_iM_f\pm1}$ factors indicate the  values of the magnetic quantum numbers for the initial state, given a fixed final state value of $M_f$. The  ``$\pm$'' sign preceding the $\gamma$ term originates from the rank-two compound tensor of the electric fields $(\hat{\v{\varepsilon}}\otimes\hat{\v{e}})^{(2)}_{M_Q}$ when the value of $M_Q$ is changed from $+1$ to $-1$.

The values of the angular factors $C^{F_fM_f}_{F_iM_f\pm1}$ and $U^{F_fM_f}_{F_iM_f\pm1}$ relevant to our computation are, explicitly
\begin{subequations}
    \begin{align}
        C^{4-4}_{3-3}&=C^{44}_{33}=-C^{33}_{44}=-C^{3-3}_{4-4}=\sqrt{7/8}\,,\\
        U^{4-4}_{3-3}&=-U^{44}_{33}=U^{33}_{44}=-U^{3-3}_{4-4}=-42\sqrt{3}\,.
    \end{align}
\end{subequations}
It is clear that these factors satisfy the following identities
\begin{subequations}
    \begin{align}\label{eq:identities}
C^{F_fM_f}_{F_iM_f\pm1}&=C^{F_f-M_f}_{F_i-M_f\mp1}\,,\\
U^{F_fM_f}_{F_iM_f\pm1}&=-U^{F_f-M_f}_{F_i-M_f\mp1}\,.
\end{align}
\end{subequations}



The measured quantities~\cite{WooBenRob99} are the transition rates $R^\pm$, Eqs.~\eqref{Eq:TheRate}, whose computation involves squaring out the sum of the Stark and PNC transition amplitudes. 
Our generalized Stark-induced amplitude is given by Eq.~\eqref{eq:revised-Stark-ind-amp}. The 
 simplified PNC~\eqref{eq:APNC} amplitude reads~\cite{WooBenRob99}
\begin{align}
A^{\mathrm{PNC}}_{i\rightarrow{f}}
=\mp\mathrm{lm}(E1_{\mathrm{PNC}})\mathcal{E}_L^{I}C^{F_fM_f}_{F_iM_f\pm{1}}\delta_{M_i,\,M_f\pm{1}}\,,
    \end{align} 
Note that while $A^{\mathrm{Stark}}_{i\rightarrow{f}}$ depends on the $z$ component of the laser field, $A^{\mathrm{PNC}}_{i\rightarrow{f}}$ depends on $\mathcal{E}_L^{I}=|\mathcal{E}_L^x|$. 
Then the generalized rates $R^+$ and $R^-$ for the two  transitions of opposite handedness are given by
\begin{subequations}\label{eq:rate-eqns-diff-handedness}
    \begin{align}
R^+ & \equiv 
R(F_i,M_f-1\rightarrow{F_f,M_f}) \nonumber\\
&=\beta^2 
(\mathcal{E}_S^x \mathcal{E}_L^z)^2
\left(C^{F_fM_f}_{F_iM_f-1}\right)^2\nonumber\\
&-\beta\gamma{\mathcal{E}_S^x \mathcal{E}_L^z}^2
C^{F_fM_f}_{F_iM_f-1}U^{F_fM_f}_{F_iM_f-1}\nonumber\\
&+2\beta{\mathrm{Im}(E1_{\rm PNC})}{\mathcal{E}_S^x}\mathcal{E}_L^z\mathcal{E}_{L}^{I}
\left(C^{F_fM_f}_{F_iM_f-1}\right)^2\nonumber\\
&-\gamma\mathcal{E}_L^z{\mathcal{E}_S^x}\mathcal{E}_{L}^{I}{\mathrm{Im}(E1_{\rm PNC})}C_{F_iM_f-1}^{F_fM_f}U^{F_fM_f}_{F_iM_f-1}\,,\label{eq:R+qe}\\
R^-&\equiv 
R(F_i,M_f+1\rightarrow{F_f,M_f}) \nonumber\\
&= \beta^2(\mathcal{E}_S^x\mathcal{E}_L^z)^2
(C^{F_fM_f}_{F_iM_f+1})^2\nonumber\\
&+\beta\gamma\left(\mathcal{E}_S^x \mathcal{E}_L^z\right)^2 \,C^{F_fM_f}_{F_iM_f+1}U^{F_fM_f}_{F_iM_f+1}\nonumber\\
&-2\beta\,{\mathrm{Im}(E1_{\rm PNC})}{\mathcal{E}_S^x}\mathcal{E}_L^z\mathcal{E}_{L}^{I}
\left(C^{F_fM_f}_{F_iM_f+1}\right)^2\nonumber\\
&-\gamma\mathcal{E}_L^z{\mathcal{E}_S^x}\mathcal{E}_{L}^{I}{\mathrm{Im}(E1_{\rm PNC})}C_{F_iM_f+1}^{F_fM_f}U^{F_fM_f}_{F_iM_f+1}\,.\label{eq:R-qe}
\end{align}
\end{subequations}
In the above expressions, $F_i$ and $F_f$ remain fixed in $R^\pm$, while the sign of $M_f$ flips when going from Eq.~\eqref{eq:R+qe} to Eq.~\eqref{eq:R-qe}. We remind the reader that we focus on the transitions between stretched hyperfine states. For example, for the $\ket{3,\,3}
\rightarrow \ket{4,\,4}$ transition\footnote{Here we used the abbreviation $\ket{F_i,\,M_i}
\rightarrow \ket{F_f,\,M_f}$, suppressing the electronic term parts of the wave-functions.}
one would use the $R^+$ expression,
while the matching transition of opposite handedness would be 
 $\ket{3,\,-3}
\rightarrow \ket{4,\,-4}$ with the $R^-$ expression to be used.
We will distinguish between four transition rates $R^+_{3\rightarrow{4}}$, $R^-_{3\rightarrow{4}}$, $R^+_{4\rightarrow{3}}$, and $R^-_{4\rightarrow{3}}$ referring to the transitions  $\ket{3,\,3}
\rightarrow \ket{4,\,4}$, $\ket{3,\,-3}\rightarrow\ket{4,\,-4}$, $\ket{4,\,-4}\rightarrow\ket{3,\,-3}$, and $\ket{4,\,4}\rightarrow\ket{3,\,3}$, respectively. For the sake of clarity, we have also suppressed the $F_i \rightarrow F_f$ superscripts in various polarizabilities.

Following Ref.~\cite{WooBenRob99}, we are interested in the rate ratio
\begin{equation}
       r_{F_i \rightarrow F_f} \equiv 
    \left( \frac{R^+-R^-}{R^++R^-} \right)_{F_i \rightarrow F_f}
\end{equation}
as it separates out the PNC amplitude. With the help of Eq.~\eqref{eq:rate-eqns-diff-handedness} and the identifies~\eqref{eq:identities}, the rate ratio generalizes to
\begin{align}\label{eq:new_ratio_of_rate}
r_{F_i \rightarrow F_f} =
\frac{1+ \frac{\gamma^{F_i \rightarrow F_f}}{2\beta^{F_i \rightarrow F_f}} \,
\frac{{U^{F_fM_f}_{F_iM_f+1}}}{{C^{F_fM_f}_{F_iM_f+1}}}
}{1+\frac{\gamma^{F_i \rightarrow F_f}}{\beta^{F_i \rightarrow F_f}} \, {\frac{U^{F_fM_f}_{F_iM_f+1}}{C^{F_fM_f}_{F_iM_f+1}}}}
\frac{{2\mathrm{Im}(E1_\mathrm{PNC}^{F_i \rightarrow F_f})}\mathcal{E}_L^{I}}{\beta^{F_i \rightarrow F_f}\mathcal{E}_L^z{\mathcal{E}_S^x}}
\,,
\end{align}
where we have emphasized the nuclear-spin dependence of the PNC amplitude by reintroducing the ${F_i \rightarrow F_f}$ superscript into $\beta$ and $\gamma$. 
In the limit of vanishing tensor polarizabilities $\gamma^{F_i \rightarrow F_f}$
and $\beta$ being independent of $F_i$ and $F_f$, Eq.~\eqref{eq:new_ratio_of_rate} reproduces the Boulder experiment's  expression~\cite{WooBenCho97,WooBenRob99}
\begin{equation}
 r_{F_i \rightarrow F_f}^\mathrm{Boulder} = 
\frac{2\, {\mathrm{Im}(E1_\mathrm{PNC}^{F_i \rightarrow F_f})}\mathcal{E}_L^{I}}{
\beta^{[2]}\mathcal{E}_L^z{\mathcal{E}_S^x}}\,.    
\end{equation}

From Eq.~\eqref{eq:new_ratio_of_rate}, the ratios $r_{F_i \rightarrow F_f}$ for the  $F_i=3\rightarrow{F_f=4}$ and $F_i=4\rightarrow{F_f=3}$ transitions are, explicitly
\begin{subequations}
\label{eq:r34andr43}
\begin{align}
r_{3\rightarrow{4}}&=\frac{2-12\sqrt{42}\frac{\gamma^{3\rightarrow{4}}}{\beta^{3\rightarrow{4}}}}{1-12\sqrt{42}\frac{\gamma^{3\rightarrow{4}}}{\beta^{3\rightarrow 4}}}\frac{{\mathrm{Im}(E1_{\rm PNC}^{3\rightarrow{4}})}\mathcal{E}^I_{L}}{\beta^{3\rightarrow 4}\mathcal{E}_L^z{\mathcal{E}_S^x}}\,,\label{eq:r34}\\
r_{4\rightarrow{3}}&=\frac{2+12\sqrt{42}\frac{\gamma^{4\rightarrow{3}}}{\beta^{4\rightarrow{3}}}}{1+12\sqrt{42}\frac{\gamma^{4\rightarrow{3}}}{\beta^{4\rightarrow{3}}}}\frac{{\mathrm{Im}(E1_{\rm PNC}^{4\rightarrow{3}})}\mathcal{E}^I_L}{\beta^{4\rightarrow{3}}\mathcal{E}_L^z{\mathcal{E}_S^x}}\,,\label{eq:r43}
\end{align}
\end{subequations}
which can be further simplified to
\begin{subequations}
\label{eq:revised-r34-r43}
    \begin{align}
    \label{eq:revised-r34}
r_{3\rightarrow4}&\approx r_{3 \rightarrow 4}^\mathrm{Boulder}\left(1- 
\frac{\delta \beta^{3\rightarrow 4}}{\beta^{[2]}}  +6
\sqrt{42}\frac{\gamma^{3\rightarrow{4}}}{\beta^{[2]}}\right)\,,\\
    \label{eq:revised-r43}
r_{4\rightarrow{3}}&\approx 
r_{4 \rightarrow 3}^\mathrm{Boulder}
\left(1
- 
\frac{\delta \beta^{4\rightarrow 3}}{\beta^{[2]}}
-6\sqrt{42}\frac{\gamma^{4\rightarrow{3}}}{\beta^{[2]}}\right)\,,
    \end{align}
\end{subequations}
where the  last two terms in the brackets are the $F$-dependent corrections to the Boulder expressions.
With our results from Table~\ref{Tab:Third-order-polarizabilities}, the corrections evaluate to $-5\times 10^{-5}$ 
and $3\times 10^{-5}$ for the $3\rightarrow4$ and the $4\rightarrow3$ transitions, respectively. These are smaller than the  experimental uncertainties in the $\mathrm{Im}({E1}_{\mathrm{PNC}}^{F_i\rightarrow{F_f}})$ determination.

The PNC amplitudes $E1_{\rm{PNC}}$ include both the 
nuclear-spin-independent and nuclear-spin-dependent contributions. The largest impact of our analysis is on the extraction of the nuclear-spin-dependent part (which includes the anapole moment contribution). If we neglect the hyperfine corrections to the transition polarizabilities,
the anapole moment contribution is extracted as~\cite{WooBenRob99}
\begin{align}\label{eq:rate_BBB}
    \frac{\mathrm{Im}(E1_{\mathrm{PNC}}^{\mathrm{anapole}})^\mathrm{Boulder}}{\beta^{[2]}}=
\left(r_{3\rightarrow4}^\mathrm{Boulder}-r_{4\rightarrow3}^\mathrm{Boulder} \right)\frac{{\mathcal{E}_S^x}\mathcal{E}_L^z}{2\mathcal{E}^I_L} \,,
\end{align}
where the authors of Ref.~\cite{WooBenRob99} associated the measured rates with $r_{F_i\rightarrow F_f}^\mathrm{Boulder}$. The measured rates $r_{F_i \rightarrow F_f}$ are, however, more accurately given as in Eq.~\eqref{eq:new_ratio_of_rate}.

To account for the nuclear-spin dependent effects on transition polarizabilities, we thus reexpress $r^{\rm Boulder}_{3\rightarrow4}$ and $r^{\rm Boulder}_{4\rightarrow3}$ in terms of $r_{3\rightarrow4}$ and $r_{4\rightarrow3}$ using Eqs.~\eqref{eq:revised-r34-r43} and use these ``adjusted'' Boulder rates in Eq.~\eqref{eq:rate_BBB}. With our semi-empirical values from Table~{\ref{Tab:Third-order-polarizabilities}}, we find $r^{\rm Boulder}_{3\rightarrow4}=1.00005\, r_{3\rightarrow 4}$ and $r^{\rm Boulder}_{4\rightarrow3}=0.99997\, r_{4\rightarrow 3}$, respectively, which cause the extracted value of ${\rm Im}(E1_{\rm{PNC}}^{4\rightarrow3})/\beta^{[2]}$ to decrease by $3\times{10^{-5}}$ while ${\rm Im}(E1_{\rm{PNC}}^{3\rightarrow4})/\beta^{[2]}$ to increase by $5\times{10^{-5}}$. Because both ${\rm Im}(E1_{\rm{PNC}}^{4\rightarrow3})/\beta^{[2]}$, and ${\rm Im}({E1}_{\rm{PNC}}^{3\rightarrow4})/\beta^{[2]}$ were reported~\cite{WooBenRob99} at about $1.6\, \rm{mV/cm}$, this means that the anapole contribution in our evaluation is slightly smaller, by about $\sim{1}\times{10^{-4}}\, \mathrm{mV/cm}$. The reported~\cite{WooBenRob99} value of the anapole moment is $0.077(11)\, \mathrm{mV/cm}$ so our correction of $1\times{10^{-4} \, \mathrm{mV/cm}}$ is below the uncertainty. This suggests that the impact due to the spin-dependent effects on polarizabilities is negligible at the current level of experimental uncertainty.

\subsection{The effect of hyperfine-mediated polarizabilities on the \texorpdfstring{$\alpha/\beta$}{} ratio analysis} 
\label{Sec:beta-alpha-ratio}
We now turn our attention to the another Boulder experiment~\cite{ChoWooBen97} which used the Stark-interference technique to determine the ratio of scalar and vector polarizabilities $\beta/\alpha$ in Cs\footnote{We note in passing that the authors of Ref.~\cite{ChoWooBen97} refer to $\beta$ as a ``tensor'' polarizability, while we call it ``vector''  to be consistent with the literature and to distinguish it from the true tensor $\gamma$ contribution.}.
This measured ratio is important in deducing the value of $\beta$ through the more computationally reliable determination of $\alpha$ (see, e.g., Ref.~\cite{TraXiaDer23-2ndOrderBeta} and the references therein). An accurate value of $\beta$ is required for extracting the PNC amplitude from the APV measurement as described in Sec.~\ref{Sec:ReinterpretationPNC}.

In the $\beta/\alpha$ experiment~\cite{ChoWooBen97}, the $^{133}\mathrm{Cs}$ atoms were spin-polarized by an external magnetic field aligned along the $y$-axis. This magnetic field defined the quantization axis that is different from that of the APV experiment described in Sec.~\ref{Sec:ReinterpretationPNC}.
To simplify our analysis, we thus define a new coordinate system $(x', y',z')$ obtained by a rotation from the $(x, y, z)$ laboratory frame defined in Sec.~\ref{Sec:ReinterpretationPNC}. The unit vectors in this new system are related to those in the frame in Sec.~\ref{Sec:ReinterpretationPNC} as follows: $\hat{\v{z}}'=\hat{\v{y}}$, $\hat{\v{y}}'=\hat{\v{x}}$, and $\hat{\v{x}}'=\hat{\v{z}}$. This transformation aligns the quantization axis with $\hat{\v{z}}'$ while preserving the handedness of the coordinate system. As a result, the electric fields in this new reference frame are given by $\v{\mathcal{E}}_S'=\mathcal{E}_S^x\hat{\v{y}}'$ and $\v{\mathcal{E}}_L'=\mathcal{E}_L^z\hat{\v{x}}'+i\mathcal{E}^I_L\hat{\v{y}}'$.

The $^{133}\mathrm{Cs}$ atoms in the $\alpha/\beta$ experiment~\cite{ChoWooBen97} underwent transitions from the initial $6S_{1/2},\,F_i=3,\,M_i=3$ state to the final  $7S_{1/2},\,F_f=3,\,M_f=3$ state. This particular choice of states guarantees an nonvanishing contribution of the scalar polarizability to the Stark-induced amplitude, Eq.~({\ref{eq:revised_amplitude}}). Then for the described experimental geometry, one has  
\begin{align}
  A_{i\rightarrow{f}}^{\rm{Stark}}=& i\alpha^{3\rightarrow{3}}\, {\mathcal{E}^I_L}\mathcal{E}_S^x+i\beta^{3\rightarrow{3}} \, {\mathcal{E}_S^x\mathcal{E}_L^z}C^{F_f,\,F_i}_{M_f,M_i} \nonumber\\
+& i K\, \gamma^{3\rightarrow{3}}\mathcal{E}_S^x\mathcal{E}^I_L\,,  \label{Eq:AStarkforaOverb}
\end{align}
where $K \equiv -i w_2( \hat{\v{\varepsilon}},\hat{\v{e}}) \, \brad{F_f=3}\{I\otimes{I}^{(2)}\}\ketd{F_i=3}$. Explicitly, since $\brad{F_f=3}\{I\otimes{I}\}^{(2)}\ketd{F_i=3}=-42\sqrt{35}$ (see Eq.~\eqref{Eq:II2numericalVals}) and $w_2( \hat{\v{\varepsilon}},\hat{\v{e}}) =- (i/6)\sqrt{5/14}$, $K =  35/\sqrt{2}$.
Here, the angular coefficient $C^{F_f\,F_i}_{M_f\,M_i}$ is defined as $C^{F_fF_i}_{M_fM_i}=g_F{\langle{M_{F}}\rangle}$ with the gyromagnetic ratio $g_F=-1/4$ and $\langle{M_F}\rangle$ being a population average over all the possible magnetic quantum numbers~\cite{ChoWooBen97}.

In contrast to the APV experiment of Sec.~\ref{Sec:ReinterpretationPNC}, the parity reversal in the $\alpha/\beta$ experiment was effected by switching the laser polarization from the left  to right 
elliptical polarization, which is equivalent to reversing the sign of the $\mathcal{E}^I_L$ in Eq.~\eqref{Eq:AStarkforaOverb}. 
This reversal flips the sign 
of the scalar and tensor contributions, while preserving the sign of the vector term in Eq.~\eqref{Eq:AStarkforaOverb}. It is clear that the interference term extracted in the experiment contains the combination 
$\left(\alpha^{3\rightarrow{3}} +K \gamma^{3\rightarrow{3}} \right)  \beta^{3\rightarrow{3}}$. This means that we need to interpret
\begin{equation} \label{Eq:aOverbReplacement}
\frac{\alpha}{\beta} \rightarrow \left(\frac{\alpha}{\beta}\right)_\mathrm{eff} = \frac{
\alpha^{3\rightarrow{3}} +K \gamma^{3\rightarrow{3}}}{\beta^{3\rightarrow{3}}}  \,,
\end{equation}
as being the ratio measured by Ref.~\cite{ChoWooBen97}.

To prove Eq.~\eqref{Eq:aOverbReplacement}, we recall that the experiment~\cite{ChoWooBen97} employed a complementary modulation of the DC electric field strength synchronous
with the elliptical polarization reversals. Two Stark-induced rates were measured, 
\begin{subequations}
    {\label{eq:rate-equations}}
\begin{align}
R^+&=\left|\alpha^{3\rightarrow{3}}\mathcal{E}^I_L+\beta^{3\rightarrow{3}}\mathcal{E}_L^zC^{F_f\,F_i}_{M_fM_i}+K 
\gamma^{3\rightarrow{3}}\mathcal{E}^I_L 
\right|^2
\left(\mathcal{E}_{S,1}^{x}\right)^{{2}}\,,\\
R^-&=\left|\alpha^{3\rightarrow{3}}\mathcal{E}^I_L-\beta^{3\rightarrow{3}}\mathcal{E}_L^zC^{F_fF_i}_{M_fM_i}+ K \gamma^{3\rightarrow{3}}\mathcal{E}^I_L\right|^2
\left(\mathcal{E}_{S,2}^{x}\right)^{2}\,,
\end{align}
\end{subequations}
where $\mathcal{E}_{S,1}^{x}$ and $\mathcal{E}_{S,2}^{x}$ stand for the magnitudes of the two DC electric fields, whereas $\mathcal{E}_L^z$ and $\mathcal{E}_{L}^{I}\equiv{\mathrm{Im}(\mathcal{E}_L^{x})}$ are the magnitudes of the two components of the laser field driving the transition. The fields $\mathcal{E}_{S,1}^{x}$ and $\mathcal{E}_{S,2}^{x}$ were adjusted until there was no modulation of the rate signal under reversals of the laser field's polarization. This amounts to equating the two rates in Eqs.~\eqref{eq:rate-equations}, thus leading to 
\begin{align}\label{eq:ratio-extraction}
\frac{\mathcal{E}_{S,2}^{x}-\mathcal{E}_{S,1}^{x}}{\mathcal{E}_{S,2}^{x}+\mathcal{E}_{S,1}^{x}}=
\frac{\beta^{3\rightarrow{3}}}{\alpha^{3\rightarrow{3}}+
K\gamma^{3\rightarrow{3}}}
\frac{
\mathcal{E}_L^z}{\mathcal{E}^I_{L}}C^{F_fF_i}_{M_fM_i}\,.
\end{align}
The inverse of the first factor on the r.h.s. of Eq.~\eqref{Eq:aOverbReplacement},
was extracted using the above equation and identified as $\alpha/\beta$ in Ref.~\cite{ChoWooBen97}. As mentioned above, the measured quantity is in fact $\left({\alpha}/{\beta}\right)_\mathrm{eff}=\left(\alpha^{3\rightarrow{3}} +K \gamma^{3\rightarrow{3}} \right) / \beta^{3\rightarrow{3}}$.

To the best of our knowledge, all the previous literature has identified  the measured~\cite{ChoWooBen97} $\alpha/\beta$ ratio with $\alpha^{[2]}/\beta^{[2]}$, neglecting the hyperfine corrections to transition polarizabilities. We  extract this ratio from the measured value~\cite{ChoWooBen97} 
$\left({\alpha}/{\beta}\right)_\mathrm{eff} = -9.905(11) $ as 
\begin{equation}\label{eq:ab=abeff}
 \frac{\alpha^{[2]}}{\beta^{[2]}} \approx 
 \left(\frac{\alpha}{\beta}\right)_\mathrm{eff}
 \left( 1 - 
 \frac{\delta\alpha^{3\rightarrow{3}}}{\alpha^{[2]}} 
 - 
 \frac{K \gamma^{3\rightarrow{3}}}{\alpha^{[2]}}
 + 
  \frac{\delta\beta^{3\rightarrow{3}}}{\beta^{[2]}} 
 \right)\,.
\end{equation}
With the recommended values of  $\alpha^{[2]}$ and $\beta^{[2]}$ as in Eqs.~\eqref{Eq:RecVals} and the hyperfine corrections from Table~\ref{Tab:Third-order-polarizabilities}, 
the corrective factor on the r.h.s of Eq.~\eqref{eq:ab=abeff} evaluates to $(1+1.3 \times 10^{-4})$, equivalent to a $\sim 0.01\%$ fractional correction to the value of $\frac{\alpha^{[2]}}{\beta^{[2]}}$. 
The inclusion of the hyperfine correction thus
modifies the last significant digit of the reported result, leading to
\begin{equation}
    \frac{\alpha^{[2]}}{\beta^{[2]}} = -9.906(11) \,,
\end{equation}
but is below the 0.1\% accuracy of the experiment~\cite{ChoWooBen97}.

\section{Conclusion}
In summary, we have introduced and evaluated the hyperfine corrections to the polarizabilities, which include the non-vanishing tensor transition polarizability $\gamma$. These HFI-mediated effects lead to a slightly smaller anapole moment extracted from the measurements of atomic parity violation by the Boulder group~\cite{WooBenCho97,WooBenRob99}. However, our computed correction is insufficient to resolve the tension with the nuclear physics interpretation and data. We also showed that the effects of the tensor transition polarizability $\gamma$ and hyperfine corrections to the scalar, $\alpha$, and vector, $\beta$, transition polarizabilities are minor but not negligible for the determination of the $\alpha/\beta$ ratio from the measurements~\cite{ChoWooBen97}. As the accuracy of experiments improves, our analysis should prove useful for interpretation of future measurements.

\section*{Acknowledgements}
This work was supported in part by the U.S. National Science Foundation grants PHY-1912465 and PHY-2207546, by the Sara Louise Hartman endowed professorship in Physics, and by the Center for Fundamental Physics at Northwestern University.

\bibliographystyle{apsrev4-2}
\bibliography{Thesis.bib}

\begin{thebibliography}{32}%
\makeatletter
\providecommand \@ifxundefined [1]{%
 \@ifx{#1\undefined}
}%
\providecommand \@ifnum [1]{%
 \ifnum #1\expandafter \@firstoftwo
 \else \expandafter \@secondoftwo
 \fi
}%
\providecommand \@ifx [1]{%
 \ifx #1\expandafter \@firstoftwo
 \else \expandafter \@secondoftwo
 \fi
}%
\providecommand \natexlab [1]{#1}%
\providecommand \enquote  [1]{``#1''}%
\providecommand \bibnamefont  [1]{#1}%
\providecommand \bibfnamefont [1]{#1}%
\providecommand \citenamefont [1]{#1}%
\providecommand \href@noop [0]{\@secondoftwo}%
\providecommand \href [0]{\begingroup \@sanitize@url \@href}%
\providecommand \@href[1]{\@@startlink{#1}\@@href}%
\providecommand \@@href[1]{\endgroup#1\@@endlink}%
\providecommand \@sanitize@url [0]{\catcode `\\12\catcode `\$12\catcode
  `\&12\catcode `\#12\catcode `\^12\catcode `\_12\catcode `\%12\relax}%
\providecommand \@@startlink[1]{}%
\providecommand \@@endlink[0]{}%
\providecommand \url  [0]{\begingroup\@sanitize@url \@url }%
\providecommand \@url [1]{\endgroup\@href {#1}{\urlprefix }}%
\providecommand \urlprefix  [0]{URL }%
\providecommand \Eprint [0]{\href }%
\providecommand \doibase [0]{https://doi.org/}%
\providecommand \selectlanguage [0]{\@gobble}%
\providecommand \bibinfo  [0]{\@secondoftwo}%
\providecommand \bibfield  [0]{\@secondoftwo}%
\providecommand \translation [1]{[#1]}%
\providecommand \BibitemOpen [0]{}%
\providecommand \bibitemStop [0]{}%
\providecommand \bibitemNoStop [0]{.\EOS\space}%
\providecommand \EOS [0]{\spacefactor3000\relax}%
\providecommand \BibitemShut  [1]{\csname bibitem#1\endcsname}%
\let\auto@bib@innerbib\@empty
\bibitem [{\citenamefont {Noecker}\ \emph {et~al.}(1988)\citenamefont
  {Noecker}, \citenamefont {Masterson},\ and\ \citenamefont
  {Wieman}}]{Noecker1988PrecisionTheory}%
  \BibitemOpen
  \bibfield  {author} {\bibinfo {author} {\bibfnamefont {M.~C.}\ \bibnamefont
  {Noecker}}, \bibinfo {author} {\bibfnamefont {B.~P.}\ \bibnamefont
  {Masterson}},\ and\ \bibinfo {author} {\bibfnamefont {C.~E.}\ \bibnamefont
  {Wieman}},\ }\href {https://doi.org/10.1103/PhysRevLett.61.310} {\bibfield
  {journal} {\bibinfo  {journal} {Phys. Rev. Lett.}\ }\textbf {\bibinfo
  {volume} {61}},\ \bibinfo {pages} {310} (\bibinfo {year} {1988})}\BibitemShut
  {NoStop}%
\bibitem [{\citenamefont {Wood}\ \emph {et~al.}(1997)\citenamefont {Wood},
  \citenamefont {Bennett}, \citenamefont {Cho}, \citenamefont {Masterson},
  \citenamefont {Roberts}, \citenamefont {Tanner},\ and\ \citenamefont
  {Wieman}}]{WooBenCho97}%
  \BibitemOpen
  \bibfield  {author} {\bibinfo {author} {\bibfnamefont {C.~S.}\ \bibnamefont
  {Wood}}, \bibinfo {author} {\bibfnamefont {S.~C.}\ \bibnamefont {Bennett}},
  \bibinfo {author} {\bibfnamefont {D.}~\bibnamefont {Cho}}, \bibinfo {author}
  {\bibfnamefont {B.~P.}\ \bibnamefont {Masterson}}, \bibinfo {author}
  {\bibfnamefont {J.~L.}\ \bibnamefont {Roberts}}, \bibinfo {author}
  {\bibfnamefont {C.~E.}\ \bibnamefont {Tanner}},\ and\ \bibinfo {author}
  {\bibfnamefont {C.~E.}\ \bibnamefont {Wieman}},\ }\href
  {https://www.science.org/doi/10.1126/science.275.5307.1759} {\bibfield
  {journal} {\bibinfo  {journal} {Science}\ }\textbf {\bibinfo {volume}
  {275}},\ \bibinfo {pages} {1759} (\bibinfo {year} {1997})}\BibitemShut
  {NoStop}%
\bibitem [{\citenamefont {Haxton}\ and\ \citenamefont
  {Wieman}(2001)}]{HaxWie01}%
  \BibitemOpen
  \bibfield  {author} {\bibinfo {author} {\bibfnamefont {W.~C.}\ \bibnamefont
  {Haxton}}\ and\ \bibinfo {author} {\bibfnamefont {C.~E.}\ \bibnamefont
  {Wieman}},\ }\href {https://doi.org/10.1146/annurev.nucl.51.101701.132458}
  {\bibfield  {journal} {\bibinfo  {journal} {Annu. Rev. Nucl. Part.}\ }\textbf
  {\bibinfo {volume} {51}},\ \bibinfo {pages} {261} (\bibinfo {year}
  {2001})}\BibitemShut {NoStop}%
\bibitem [{\citenamefont {Desplanques}\ \emph {et~al.}(1980)\citenamefont
  {Desplanques}, \citenamefont {Donoghue},\ and\ \citenamefont
  {Holstein}}]{DDH1980}%
  \BibitemOpen
  \bibfield  {author} {\bibinfo {author} {\bibfnamefont {B.}~\bibnamefont
  {Desplanques}}, \bibinfo {author} {\bibfnamefont {J.~F.}\ \bibnamefont
  {Donoghue}},\ and\ \bibinfo {author} {\bibfnamefont {B.~R.}\ \bibnamefont
  {Holstein}},\ }\href
  {https://doi.org/http://dx.doi.org/10.1016/0003-4916(80)90217-1} {\bibfield
  {journal} {\bibinfo  {journal} {Ann. Phys.}\ }\textbf {\bibinfo {volume}
  {124}},\ \bibinfo {pages} {449} (\bibinfo {year} {1980})}\BibitemShut
  {NoStop}%
\bibitem [{\citenamefont {Bouchiat}\ and\ \citenamefont
  {Bouchiat}(1975)}]{BouBou75}%
  \BibitemOpen
  \bibfield  {author} {\bibinfo {author} {\bibfnamefont {M.~A.}\ \bibnamefont
  {Bouchiat}}\ and\ \bibinfo {author} {\bibfnamefont {C.}~\bibnamefont
  {Bouchiat}},\ }\href
  {https://jphys.journaldephysique.org/articles/jphys/abs/1975/06/jphys_1975__36_6_493_0/jphys_1975__36_6_493_0.html}
  {\bibfield  {journal} {\bibinfo  {journal} {J. Phys.}\ }\textbf {\bibinfo
  {volume} {36}},\ \bibinfo {pages} {493} (\bibinfo {year} {1975})}\BibitemShut
  {NoStop}%
\bibitem [{\citenamefont {Tan}\ \emph {et~al.}(2023)\citenamefont {Tan},
  \citenamefont {Xiao},\ and\ \citenamefont
  {Derevianko}}]{TraXiaDer23-2ndOrderBeta}%
  \BibitemOpen
  \bibfield  {author} {\bibinfo {author} {\bibfnamefont {H.~B.~T.}\
  \bibnamefont {Tan}}, \bibinfo {author} {\bibfnamefont {D.}~\bibnamefont
  {Xiao}},\ and\ \bibinfo {author} {\bibfnamefont {A.}~\bibnamefont
  {Derevianko}},\ }\href@noop {} {} (\bibinfo {year} {2023}),\ \Eprint
  {https://arxiv.org/abs/2306.09573} {arXiv:2306.09573 [physics.atom-ph]}
  \BibitemShut {NoStop}%
\bibitem [{\citenamefont {Varshalovich}\ \emph {et~al.}(1988)\citenamefont
  {Varshalovich}, \citenamefont {Moskalev},\ and\ \citenamefont
  {Khersonskii}}]{VarMosKhe88}%
  \BibitemOpen
  \bibfield  {author} {\bibinfo {author} {\bibfnamefont {D.~A.}\ \bibnamefont
  {Varshalovich}}, \bibinfo {author} {\bibfnamefont {A.~N.}\ \bibnamefont
  {Moskalev}},\ and\ \bibinfo {author} {\bibfnamefont {V.~K.}\ \bibnamefont
  {Khersonskii}},\ }\href
  {https://www.worldscientific.com/worldscibooks/10.1142/0270#t=aboutBook}
  {\emph {\bibinfo {title} {{Quantum Theory of Angular Momentum}}}}\ (\bibinfo
  {publisher} {World Scientific},\ \bibinfo {address} {Singapore},\ \bibinfo
  {year} {1988})\BibitemShut {NoStop}%
\bibitem [{\citenamefont {Dzuba}\ \emph {et~al.}(1997)\citenamefont {Dzuba},
  \citenamefont {Flambaum},\ and\ \citenamefont {Sushkov}}]{Dzuba1997}%
  \BibitemOpen
  \bibfield  {author} {\bibinfo {author} {\bibfnamefont {V.~A.}\ \bibnamefont
  {Dzuba}}, \bibinfo {author} {\bibfnamefont {V.~V.}\ \bibnamefont
  {Flambaum}},\ and\ \bibinfo {author} {\bibfnamefont {O.~P.}\ \bibnamefont
  {Sushkov}},\ }\href {https://doi.org/10.1103/PhysRevA.56.R4357} {\bibfield
  {journal} {\bibinfo  {journal} {Phys. Rev. A}\ }\textbf {\bibinfo {volume}
  {56}},\ \bibinfo {pages} {R4357} (\bibinfo {year} {1997})}\BibitemShut
  {NoStop}%
\bibitem [{\citenamefont {Safronova}\ \emph {et~al.}(1999)\citenamefont
  {Safronova}, \citenamefont {Johnson},\ and\ \citenamefont
  {Derevianko}}]{Safronova1999}%
  \BibitemOpen
  \bibfield  {author} {\bibinfo {author} {\bibfnamefont {M.~S.}\ \bibnamefont
  {Safronova}}, \bibinfo {author} {\bibfnamefont {W.~R.}\ \bibnamefont
  {Johnson}},\ and\ \bibinfo {author} {\bibfnamefont {A.}~\bibnamefont
  {Derevianko}},\ }\href {https://doi.org/10.1103/PhysRevA.60.4476} {\bibfield
  {journal} {\bibinfo  {journal} {Phys. Rev. A}\ }\textbf {\bibinfo {volume}
  {60}},\ \bibinfo {pages} {4476} (\bibinfo {year} {1999})}\BibitemShut
  {NoStop}%
\bibitem [{\citenamefont {Vasilyev}\ \emph {et~al.}(2002)\citenamefont
  {Vasilyev}, \citenamefont {Savukov}, \citenamefont {Safronova},\ and\
  \citenamefont {Berry}}]{Vas2002}%
  \BibitemOpen
  \bibfield  {author} {\bibinfo {author} {\bibfnamefont {A.~A.}\ \bibnamefont
  {Vasilyev}}, \bibinfo {author} {\bibfnamefont {I.~M.}\ \bibnamefont
  {Savukov}}, \bibinfo {author} {\bibfnamefont {M.~S.}\ \bibnamefont
  {Safronova}},\ and\ \bibinfo {author} {\bibfnamefont {H.~G.}\ \bibnamefont
  {Berry}},\ }\href {https://doi.org/10.1103/PhysRevA.66.020101} {\bibfield
  {journal} {\bibinfo  {journal} {Phys. Rev. A}\ }\textbf {\bibinfo {volume}
  {66}},\ \bibinfo {pages} {020101(R)} (\bibinfo {year} {2002})}\BibitemShut
  {NoStop}%
\bibitem [{\citenamefont {Dzuba}\ \emph {et~al.}(2002)\citenamefont {Dzuba},
  \citenamefont {Flambaum},\ and\ \citenamefont {Ginges}}]{Dzuba2002}%
  \BibitemOpen
  \bibfield  {author} {\bibinfo {author} {\bibfnamefont {V.~A.}\ \bibnamefont
  {Dzuba}}, \bibinfo {author} {\bibfnamefont {V.~V.}\ \bibnamefont
  {Flambaum}},\ and\ \bibinfo {author} {\bibfnamefont {J.~S.~M.}\ \bibnamefont
  {Ginges}},\ }\href {https://doi.org/10.1103/PhysRevD.66.076013} {\bibfield
  {journal} {\bibinfo  {journal} {Phys. Rev. D}\ }\textbf {\bibinfo {volume}
  {66}},\ \bibinfo {pages} {076013} (\bibinfo {year} {2002})}\BibitemShut
  {NoStop}%
\bibitem [{\citenamefont {Toh}\ \emph {et~al.}(2019)\citenamefont {Toh},
  \citenamefont {Damitz}, \citenamefont {Tanner}, \citenamefont {Johnson},\
  and\ \citenamefont {Elliott}}]{Toh2019}%
  \BibitemOpen
  \bibfield  {author} {\bibinfo {author} {\bibfnamefont {G.}~\bibnamefont
  {Toh}}, \bibinfo {author} {\bibfnamefont {A.}~\bibnamefont {Damitz}},
  \bibinfo {author} {\bibfnamefont {C.~E.}\ \bibnamefont {Tanner}}, \bibinfo
  {author} {\bibfnamefont {W.~R.}\ \bibnamefont {Johnson}},\ and\ \bibinfo
  {author} {\bibfnamefont {D.~S.}\ \bibnamefont {Elliott}},\ }\href
  {https://doi.org/10.1103/PhysRevLett.123.073002} {\bibfield  {journal}
  {\bibinfo  {journal} {Phys. \ Rev. \ Lett.}\ }\textbf {\bibinfo {volume}
  {123}},\ \bibinfo {pages} {073002} (\bibinfo {year} {2019})}\BibitemShut
  {NoStop}%
\bibitem [{\citenamefont {Bennett}\ and\ \citenamefont
  {Wieman}(1999)}]{Ben1999}%
  \BibitemOpen
  \bibfield  {author} {\bibinfo {author} {\bibfnamefont {S.~C.}\ \bibnamefont
  {Bennett}}\ and\ \bibinfo {author} {\bibfnamefont {C.~E.}\ \bibnamefont
  {Wieman}},\ }\href {https://doi.org/10.1103/PhysRevLett.82.2484} {\bibfield
  {journal} {\bibinfo  {journal} {Phys. Rev. Lett.}\ }\textbf {\bibinfo
  {volume} {82}},\ \bibinfo {pages} {2484} (\bibinfo {year}
  {1999})}\BibitemShut {NoStop}%
\bibitem [{\citenamefont {Dzuba}\ and\ \citenamefont
  {Flambaum}(2000)}]{DzuFla00}%
  \BibitemOpen
  \bibfield  {author} {\bibinfo {author} {\bibfnamefont {V.~A.}\ \bibnamefont
  {Dzuba}}\ and\ \bibinfo {author} {\bibfnamefont {V.~V.}\ \bibnamefont
  {Flambaum}},\ }\href
  {https://journals.aps.org/pra/abstract/10.1103/PhysRevA.62.052101} {\bibfield
   {journal} {\bibinfo  {journal} {Phys.\ Rev.\ A}\ }\textbf {\bibinfo {volume}
  {62}},\ \bibinfo {pages} {052101} (\bibinfo {year} {2000})}\BibitemShut
  {NoStop}%
\bibitem [{\citenamefont {Sahoo}\ and\ \citenamefont {Das}(2020)}]{sahoo20}%
  \BibitemOpen
  \bibfield  {author} {\bibinfo {author} {\bibfnamefont {B.~K.}\ \bibnamefont
  {Sahoo}}\ and\ \bibinfo {author} {\bibfnamefont {B.~P.}\ \bibnamefont
  {Das}},\ }\href@noop {} {} (\bibinfo {year} {2020}),\ \Eprint
  {https://arxiv.org/abs/2008.08941} {arXiv:2008.08941 [hep-ph]} \BibitemShut
  {NoStop}%
\bibitem [{\citenamefont {Beloy}\ \emph {et~al.}(2006)\citenamefont {Beloy},
  \citenamefont {Safronova},\ and\ \citenamefont {Derevianko}}]{BelSafDer06}%
  \BibitemOpen
  \bibfield  {author} {\bibinfo {author} {\bibfnamefont {K.}~\bibnamefont
  {Beloy}}, \bibinfo {author} {\bibfnamefont {U.~I.}\ \bibnamefont
  {Safronova}},\ and\ \bibinfo {author} {\bibfnamefont {A.}~\bibnamefont
  {Derevianko}},\ }\href {https://doi.org/10.1103/PhysRevLett.97.040801}
  {\bibfield  {journal} {\bibinfo  {journal} {Phys. Rev. Lett.}\ }\textbf
  {\bibinfo {volume} {97}},\ \bibinfo {pages} {040801} (\bibinfo {year}
  {2006})}\BibitemShut {NoStop}%
\bibitem [{\citenamefont {Dzuba}\ \emph {et~al.}(2010)\citenamefont {Dzuba},
  \citenamefont {Flambaum}, \citenamefont {Beloy},\ and\ \citenamefont
  {Derevianko}}]{DzuFlaBel10}%
  \BibitemOpen
  \bibfield  {author} {\bibinfo {author} {\bibfnamefont {V.~A.}\ \bibnamefont
  {Dzuba}}, \bibinfo {author} {\bibfnamefont {V.~V.}\ \bibnamefont {Flambaum}},
  \bibinfo {author} {\bibfnamefont {K.}~\bibnamefont {Beloy}},\ and\ \bibinfo
  {author} {\bibfnamefont {A.}~\bibnamefont {Derevianko}},\ }\href
  {https://doi.org/10.1103/PhysRevA.82.062513} {\bibfield  {journal} {\bibinfo
  {journal} {Phys. Rev. A}\ }\textbf {\bibinfo {volume} {82}},\ \bibinfo
  {pages} {062513} (\bibinfo {year} {2010})}\BibitemShut {NoStop}%
\bibitem [{\citenamefont {Rosenbusch}\ \emph {et~al.}(2009)\citenamefont
  {Rosenbusch}, \citenamefont {Ghezali}, \citenamefont {Dzuba}, \citenamefont
  {Flambaum}, \citenamefont {Beloy},\ and\ \citenamefont
  {Derevianko}}]{RosGheDzu09}%
  \BibitemOpen
  \bibfield  {author} {\bibinfo {author} {\bibfnamefont {P.}~\bibnamefont
  {Rosenbusch}}, \bibinfo {author} {\bibfnamefont {S.}~\bibnamefont {Ghezali}},
  \bibinfo {author} {\bibfnamefont {V.~A.}\ \bibnamefont {Dzuba}}, \bibinfo
  {author} {\bibfnamefont {V.~V.}\ \bibnamefont {Flambaum}}, \bibinfo {author}
  {\bibfnamefont {K.}~\bibnamefont {Beloy}},\ and\ \bibinfo {author}
  {\bibfnamefont {A.}~\bibnamefont {Derevianko}},\ }\href
  {https://doi.org/10.1103/PhysRevA.79.013404} {\bibfield  {journal} {\bibinfo
  {journal} {Phys. Rev. A}\ }\textbf {\bibinfo {volume} {79}},\ \bibinfo
  {pages} {013404} (\bibinfo {year} {2009})}\BibitemShut {NoStop}%
\bibitem [{\citenamefont {DeMille}\ and\ \citenamefont
  {Kozlov}(1998)}]{DeMille1998}%
  \BibitemOpen
  \bibfield  {author} {\bibinfo {author} {\bibfnamefont {D.}~\bibnamefont
  {DeMille}}\ and\ \bibinfo {author} {\bibfnamefont {M.}~\bibnamefont
  {Kozlov}},\ }\href@noop {} {} (\bibinfo {year} {1998}),\ \Eprint
  {https://arxiv.org/abs/physics/9801034} {arXiv:physics/9801034
  [physics.atom-ph]} \BibitemShut {NoStop}%
\bibitem [{\citenamefont {Johnson}(2007)}]{Joh07}%
  \BibitemOpen
  \bibfield  {author} {\bibinfo {author} {\bibfnamefont {W.~R.}\ \bibnamefont
  {Johnson}},\ }\href
  {https://link.springer.com/book/10.1007/978-3-540-68013-0} {\emph {\bibinfo
  {title} {{Atomic Structure Theory: Lectures on Atomic Physics}}}}\ (\bibinfo
  {publisher} {Springer},\ \bibinfo {address} {New York, NY},\ \bibinfo {year}
  {2007})\BibitemShut {NoStop}%
\bibitem [{\citenamefont {Gerginov}\ \emph {et~al.}(2003)\citenamefont
  {Gerginov}, \citenamefont {Derevianko},\ and\ \citenamefont
  {Tanner}}]{GerDerTan03}%
  \BibitemOpen
  \bibfield  {author} {\bibinfo {author} {\bibfnamefont {V.}~\bibnamefont
  {Gerginov}}, \bibinfo {author} {\bibfnamefont {A.}~\bibnamefont
  {Derevianko}},\ and\ \bibinfo {author} {\bibfnamefont {C.~E.}\ \bibnamefont
  {Tanner}},\ }\href {https://doi.org/10.1103/PhysRevLett.91.072501} {\bibfield
   {journal} {\bibinfo  {journal} {Phys. Rev. Lett.}\ }\textbf {\bibinfo
  {volume} {91}},\ \bibinfo {pages} {072501} (\bibinfo {year}
  {2003})}\BibitemShut {NoStop}%
\bibitem [{\citenamefont {Das}\ and\ \citenamefont
  {Natarajan}(2008)}]{DasNat08}%
  \BibitemOpen
  \bibfield  {author} {\bibinfo {author} {\bibfnamefont {D.}~\bibnamefont
  {Das}}\ and\ \bibinfo {author} {\bibfnamefont {V.}~\bibnamefont
  {Natarajan}},\ }\href {https://doi.org/10.1088/0953-4075/41/3/035001}
  {\bibfield  {journal} {\bibinfo  {journal} {Journal of Physics B: Atomic,
  Molecular and Optical Physics}\ }\textbf {\bibinfo {volume} {41}},\ \bibinfo
  {pages} {035001} (\bibinfo {year} {2008})}\BibitemShut {NoStop}%
\bibitem [{\citenamefont {Tran~Tan}\ and\ \citenamefont
  {Derevianko}(2023)}]{tan2023precision}%
  \BibitemOpen
  \bibfield  {author} {\bibinfo {author} {\bibfnamefont {H.~B.}\ \bibnamefont
  {Tran~Tan}}\ and\ \bibinfo {author} {\bibfnamefont {A.}~\bibnamefont
  {Derevianko}},\ }\href {https://doi.org/10.1103/PhysRevA.107.042809}
  {\bibfield  {journal} {\bibinfo  {journal} {Phys. Rev. A}\ }\textbf {\bibinfo
  {volume} {107}},\ \bibinfo {pages} {042809} (\bibinfo {year}
  {2023})}\BibitemShut {NoStop}%
\bibitem [{\citenamefont {Beloy}\ and\ \citenamefont
  {Derevianko}(2008)}]{BelDer08.DKB}%
  \BibitemOpen
  \bibfield  {author} {\bibinfo {author} {\bibfnamefont {K.}~\bibnamefont
  {Beloy}}\ and\ \bibinfo {author} {\bibfnamefont {A.}~\bibnamefont
  {Derevianko}},\ }\href {https://doi.org/10.1016/j.cpc.2008.03.004} {\bibfield
   {journal} {\bibinfo  {journal} {Comp. Phys. Comm.}\ }\textbf {\bibinfo
  {volume} {179}},\ \bibinfo {pages} {310} (\bibinfo {year}
  {2008})}\BibitemShut {NoStop}%
\bibitem [{\citenamefont {Kramida}\ \emph {et~al.}(2020)\citenamefont
  {Kramida}, \citenamefont {Ralchenko},\ and\ \citenamefont
  {Reader}}]{NIST_ASD}%
  \BibitemOpen
  \bibfield  {author} {\bibinfo {author} {\bibfnamefont {A.}~\bibnamefont
  {Kramida}}, \bibinfo {author} {\bibfnamefont {Y.}~\bibnamefont {Ralchenko}},\
  and\ \bibinfo {author} {\bibfnamefont {J.}~\bibnamefont {Reader}},\ }\href
  {https://physics.nist.gov/asd} {\bibfield  {journal} {\bibinfo  {journal}
  {NIST Atomic Spectra Database}\ } (\bibinfo {year} {2020})}\BibitemShut
  {NoStop}%
\bibitem [{\citenamefont {Belin}\ \emph {et~al.}(1976)\citenamefont {Belin},
  \citenamefont {Holmgren},\ and\ \citenamefont {Svanberg}}]{Belin1976}%
  \BibitemOpen
  \bibfield  {author} {\bibinfo {author} {\bibfnamefont {G.}~\bibnamefont
  {Belin}}, \bibinfo {author} {\bibfnamefont {L.}~\bibnamefont {Holmgren}},\
  and\ \bibinfo {author} {\bibfnamefont {S.}~\bibnamefont {Svanberg}},\ }\href
  {https://doi.org/10.1088/0031-8949/14/1-2/008} {\bibfield  {journal}
  {\bibinfo  {journal} {Phys. Scr.}\ }\textbf {\bibinfo {volume} {14}},\
  \bibinfo {pages} {39} (\bibinfo {year} {1976})}\BibitemShut {NoStop}%
\bibitem [{\citenamefont {Arimondo}\ \emph {et~al.}(1977)\citenamefont
  {Arimondo}, \citenamefont {Inguscio},\ and\ \citenamefont
  {Violino}}]{AriIngVio77}%
  \BibitemOpen
  \bibfield  {author} {\bibinfo {author} {\bibfnamefont {E.}~\bibnamefont
  {Arimondo}}, \bibinfo {author} {\bibfnamefont {M.}~\bibnamefont {Inguscio}},\
  and\ \bibinfo {author} {\bibfnamefont {P.}~\bibnamefont {Violino}},\ }\href
  {https://doi.org/10.1103/RevModPhys.49.31} {\bibfield  {journal} {\bibinfo
  {journal} {Rev. Mod. Phys.}\ }\textbf {\bibinfo {volume} {49}},\ \bibinfo
  {pages} {31} (\bibinfo {year} {1977})}\BibitemShut {NoStop}%
\bibitem [{\citenamefont {Auzinsh}\ \emph {et~al.}(2007)\citenamefont
  {Auzinsh}, \citenamefont {Bluss}, \citenamefont {Ferber}, \citenamefont
  {Gahbauer}, \citenamefont {Jarmola}, \citenamefont {Safronova}, \citenamefont
  {Safronova},\ and\ \citenamefont {Tamanis}}]{Auzinsh2007}%
  \BibitemOpen
  \bibfield  {author} {\bibinfo {author} {\bibfnamefont {M.}~\bibnamefont
  {Auzinsh}}, \bibinfo {author} {\bibfnamefont {K.}~\bibnamefont {Bluss}},
  \bibinfo {author} {\bibfnamefont {R.}~\bibnamefont {Ferber}}, \bibinfo
  {author} {\bibfnamefont {F.}~\bibnamefont {Gahbauer}}, \bibinfo {author}
  {\bibfnamefont {A.}~\bibnamefont {Jarmola}}, \bibinfo {author} {\bibfnamefont
  {M.~S.}\ \bibnamefont {Safronova}}, \bibinfo {author} {\bibfnamefont {U.~I.}\
  \bibnamefont {Safronova}},\ and\ \bibinfo {author} {\bibfnamefont
  {M.}~\bibnamefont {Tamanis}},\ }\href
  {https://doi.org/10.1103/PhysRevA.75.022502} {\bibfield  {journal} {\bibinfo
  {journal} {Phys. Rev. A}\ }\textbf {\bibinfo {volume} {75}},\ \bibinfo
  {pages} {022502} (\bibinfo {year} {2007})}\BibitemShut {NoStop}%
\bibitem [{\citenamefont {Derevianko}\ \emph {et~al.}(1999)\citenamefont
  {Derevianko}, \citenamefont {Safronova},\ and\ \citenamefont
  {Johnson}}]{Derevianko1999b}%
  \BibitemOpen
  \bibfield  {author} {\bibinfo {author} {\bibfnamefont {A.}~\bibnamefont
  {Derevianko}}, \bibinfo {author} {\bibfnamefont {M.~S.}\ \bibnamefont
  {Safronova}},\ and\ \bibinfo {author} {\bibfnamefont {W.~R.}\ \bibnamefont
  {Johnson}},\ }\href {https://doi.org/10.1103/PhysRevA.60.R1741} {\bibfield
  {journal} {\bibinfo  {journal} {Phys. Rev. A}\ }\textbf {\bibinfo {volume}
  {60}},\ \bibinfo {pages} {R1741} (\bibinfo {year} {1999})}\BibitemShut
  {NoStop}%
\bibitem [{\citenamefont {Cho}\ \emph {et~al.}(1997)\citenamefont {Cho},
  \citenamefont {Wood}, \citenamefont {Bennett}, \citenamefont {Roberts},\ and\
  \citenamefont {Wieman}}]{ChoWooBen97}%
  \BibitemOpen
  \bibfield  {author} {\bibinfo {author} {\bibfnamefont {D.}~\bibnamefont
  {Cho}}, \bibinfo {author} {\bibfnamefont {C.~S.}\ \bibnamefont {Wood}},
  \bibinfo {author} {\bibfnamefont {S.~C.}\ \bibnamefont {Bennett}}, \bibinfo
  {author} {\bibfnamefont {J.~L.}\ \bibnamefont {Roberts}},\ and\ \bibinfo
  {author} {\bibfnamefont {C.~E.}\ \bibnamefont {Wieman}},\ }\href
  {https://journals.aps.org/pra/abstract/10.1103/PhysRevA.55.1007} {\bibfield
  {journal} {\bibinfo  {journal} {Phys. Rev. A}\ }\textbf {\bibinfo {volume}
  {55}},\ \bibinfo {pages} {1007} (\bibinfo {year} {1997})}\BibitemShut
  {NoStop}%
\bibitem [{\citenamefont {Wood}\ \emph {et~al.}(1999)\citenamefont {Wood},
  \citenamefont {Bennett}, \citenamefont {Roberts}, \citenamefont {Cho},\ and\
  \citenamefont {Wieman}}]{WooBenRob99}%
  \BibitemOpen
  \bibfield  {author} {\bibinfo {author} {\bibfnamefont {C.~S.}\ \bibnamefont
  {Wood}}, \bibinfo {author} {\bibfnamefont {S.~C.}\ \bibnamefont {Bennett}},
  \bibinfo {author} {\bibfnamefont {J.~L.}\ \bibnamefont {Roberts}}, \bibinfo
  {author} {\bibfnamefont {D.}~\bibnamefont {Cho}},\ and\ \bibinfo {author}
  {\bibfnamefont {C.~E.}\ \bibnamefont {Wieman}},\ }\href
  {https://cdnsciencepub.com/doi/10.1139/p99-002} {\bibfield  {journal}
  {\bibinfo  {journal} {Can. J. Phys.}\ }\textbf {\bibinfo {volume} {77}},\
  \bibinfo {pages} {7} (\bibinfo {year} {1999})}\BibitemShut {NoStop}%
\bibitem [{\citenamefont {Gilbert}\ and\ \citenamefont
  {Wieman}(1986)}]{Gilbert1986}%
  \BibitemOpen
  \bibfield  {author} {\bibinfo {author} {\bibfnamefont {S.~L.}\ \bibnamefont
  {Gilbert}}\ and\ \bibinfo {author} {\bibfnamefont {C.~E.}\ \bibnamefont
  {Wieman}},\ }\href {https://doi.org/10.1103/PhysRevA.34.792} {\bibfield
  {journal} {\bibinfo  {journal} {Phys. Rev. A}\ }\textbf {\bibinfo {volume}
  {34}},\ \bibinfo {pages} {792} (\bibinfo {year} {1986})}\BibitemShut
  {NoStop}%
\end{thebibliography}%
\end{document}